# Intercalation of functional materials with phase transitions for neuromorphic applications

Xin He,[1, 2] Hua Wang,[1] Jian Sun,[3] Xixiang Zhang,[4] Kai Chang,[1] and Fei Xue[1, 2*]

[1]Center for Quantum Matter, School of Physics, Zhejiang University, Hangzhou 310027, China.

[2]College of Integrated Circuits, ZJU-Hangzhou Global Scientific and Technological Innovation Center, Zhejiang University, Hangzhou 311215, China.

[3]School of Physics, Central South University, Changsha 410083, China.

[4]Physical Science and Engineering Division, King Abdullah University of Science and Technology, Thuwal 23955-6900, Saudi Arabia.

*Correspondence: xuef@zju.edu.cn

## Abstract

The introduction of foreign ions, atoms, or molecules into emerging functional materials is crucial for manipulating material physical properties and innovating device applications. The intercalation of emerging new materials can induce multiple intrinsic changes, such as charge doping, chemical bonding, and lattice expansion, which facilitates the exploration of structural phase transformations, the tuning of symmetry-breaking-related physics, and the creation of brain-inspired advanced devices. Moreover, incorporating various hosts and intercalants enables a series of crystal structures with a rich spectrum of characteristics, greatly expanding the scope and fundamental understanding of existing materials. Herein, we summarize typically used methods for the intercalation of functional materials. We highlight recent progress in intercalation-based phase transitions and their emerging physics, i.e., ferroelectric, magnetic, insulator-metal, superconducting, and charge-density-wave phase transitions. We discuss prospective device applications for intercalation-based phase transitions, i.e., neuromorphic devices. Finally, we provide potential future research lines for promoting its further development.

## Introduction

Intercalation is a technique of inserting foreign atoms/molecules/ions into host materials, which was first developed nearly two centuries ago[1] and is blooming in the new century with the rise of two-dimensional (2D) layered materials.[2] The field of functional material intercalation has attracted considerable attention due to its potential to greatly expand the existing material family,[3-5] facilitate material preparation on the atomic level,[6-9] and tailor the physical and chemical properties of materials.[10-14] In addition to these fundamental advances, the intercalation of functional materials also enables the development of high-performance energy storage.[15-18]

More specifically, once a host is intercalated with different guests, a series of new compounds with different structures, components, and properties can be produced by design.[19,20] The large numbers of hosts and guests significantly enrich the material library. The intercalation of large ions like $SO_4^{2-}$, $BF_4^-$, and $R_4N^+$ into bulk materials can enable the preparation of atomically thin materials by expanding their interlayer spacings and further separating the neighboring layers.[21,22] In comparison to other exfoliation methods, intercalation is a more efficient and scalable approach for thinning layered materials. The intercalation of different guests can modify the physical behaviors of hosts, such as electronic,[23-27] magnetic,[28-35] optical,[36-38] optoelectronic,[39-41] thermal,[42-44] and thermoelectric[45-47] properties. Interestingly, the intercalation of certain foreign species can result in the coupling of two or more characteristics, such as the synchronous alteration of conductivity and optical transmittance,[48,49] or the coexistence of superconductivity and magnetism.[50,51] which inspires the fabrication of multifunctional electronic devices. Lastly, intercalation in the field of energy storage (e.g., Li-ion battery) has greatly revolutionized our world through its incorporation in commercial products, such as portable electronic devices (e.g., cellphones and laptops) and electronic vehicles.

One of the remarkable dynamics for intercalating functional materials is the phase transition, which typically involves ferroelectric,[52-54] magnetic,[29,34,35,50,55-58] insulator-metal (I-M),[25-27,29,57,59-63] superconducting,[50,64-67] and charge-density-wave (CDW) phase transitions.[68-72] Pioneering studies can be traced back to approximately half a century ago. In the 1960s and 1970s, alkali metal ions, transition metal ions, and molecules were intercalated into graphite[55,64] and transition metal dichalcogenides (TMDs).[56,61,62,65,68] Consequently, several exotic physical phenomena, such as superconductivity,[64,65] magnetism,[55,56] and CDW[68] were observed, thereby corroborating the efficacy of functional material intercalation. Over the following three decades, the research focused on the intercalation of bulk materials, and notable progress was rare, likely due to the lack of suitable material systems. Since the rise of atomic 2D layered materials in 2004,[73] intercalation has been promptly introduced as a means of inducing precise phase transitions on the 2D atomic scale. In 2010, the intercalation of Cu into a topological insulator, $Bi_2Se_3$, resulted in the emergence of superconductivity, providing a new strategy to manipulate the topological surface state instead of proximity to a superconductor.[66] Following substantial attempts to manipulate intercalation, CDWs were observed unambiguously in graphitic materials for the first time in 2011.[69] Surprisingly, Li et al. discovered that intercalated $SnSe_2$−$Co(Cp)_2$ superlattices, formed by non-ferromagnetic and non-superconducting building blocks, displayed the first freestanding coexistence of ferromagnetism and superconductivity in 2017.[50] In the same year, $O^{2-}$ and $H^+$ ions were intercalated into $SrCoO_{3-\delta}$ thereby inducing unprecedented tri-state phase transitions,[57] which opens the door to multifunctional devices. In the following years, protons have been intercalated into other magnetic oxides and layered materials, inducing diverse magnetic phase transitions[29,34,35] and spurring extensive applications of protonic gating. Recently, non-ferroelectric 2H-$MoS_2$ has been transformed into ferroelectric 1T''-$MoS_2$ via Li intercalation,[52] demonstrating a breakthrough in the phase engineering of non-ferroelectric to ferroelectric materials. Motivated by the previous studies in proton-driven phase transitions, our group successfully demonstrated ionic ferroelectric phase transitions for the first time by intercalating protons into the ferroelectric α-$in_2Se_3$ in 2023,[54] which could be used in creating multilevel memory devices. Surprisingly, via the mediation of the photoredox reaction, Li-intercalation-based phase transition can be greatly accelerated up to six orders of magnitude,[63] which facilitates the in-situ observation of intercalation dynamical processes. The key landmarks for the investigation of intercalation-based phase transitions are shown in Figure 1.

Intercalation-based phase transitions have been extensively utilized in energy-efficient neuromorphic devices to emulate biological functions and perform complex brain-inspired computing. Proton is one of the typical intercalants to render phase transitions in many electrochemical transistors, particularly those with metal oxide-based channels, such as $VO_2$,[74,75] $MoO_3$,[76] $WO_3$,[77] and $NdNiO_3$.[78] For instance, by altering the gate voltage of an electrochemical transistor with a $NdNiO_3$ channel, the phase across the channel can transition between the metallic $NdNiO_3$ and insulating H-$NdNiO_3$ in a reversible manner, accompanied by a corresponding change in conductance.[78] After modulating the quantity of the intercalated protons, nonvolatile multilevel conductance states are achieved via a gate pulse of down to 50 mV with an energy consumption as low as ~ 1pJ.[78] Furthermore, the lithium ion is another frequently utilized intercalant to initiate phase transitions for neuromorphic applications. For example, when $Li^+$ ions are intercalated into the $Nb_2O_5$ channel, $Nb^{5+}$ is reduced to $Nb^{4+}$ electrochemically, subsequently leading to the formation of a novel phase ($Li_xNb_2O_5$) with enhanced conductivity.[79]. Such Li-intercalation-based electrochemical transistors have been paired with conventional transistors to realize neuromorphic computing systems.[79] The energy efficiency of the one-transistor-one-electrochemical transistor-based spiking neural networks has been enhanced by approximately 30 times compared to that of the prevalent rate coding scheme.[79] In other non-transistor neuromorphic devices (i.e., multiterminal memristive devices), $Li^+$ ions have been inserted into $MoS_2$ channels, where $MoS_2$ undergoes a phase transition from the semiconducting 2H phase to the metallic 1T' phase.[80] Such $MoS_2$ devices not only display essential memristive characteristics but can also emulate complicated synaptic competition and cooperation effects observed in biology, which enables the first direct emulation of biological synaptic dynamics using intercalation-based multiple coplanar devices.[80]

In addition, significant theoretical advancements have been made in this field recently. These include theoretical exploration of intercalation chemistry in TMDs,[81] high-throughput prediction of intercalated multiferroic materials[82] and self-intercalated 2D materials with I-M phase transitions,[83] and intercalation-based coexisting superconductivity and CDW,[84] Furthermore, theoretical studies have been conducted on exploring promising TMDs for battery applications.[85] These theoretical works could offer valuable guidance for intercalation-based experimental explorations, facilitating more efficient and targeted experimentation.

In this review, we focus on the interaction-based phase transitions and related neuromorphic applications. We discuss four typical intercalation methods, including liquid-phase intercalation, electrochemical intercalation, vapor-phase intercalation, and solid-phase intercalation. We briefly introduce a number of important characterization techniques used to examine the phase transitions, such as transmission electron microscopy (TEM), scanning tunneling microscopy (STM), atomic force microscopy (AFM), X-ray diffraction (XRD), X-ray photoelectron spectroscopy (XPS), and Raman spectroscopy. We then present a comprehensive overview of the intercalation-based ferroelectric, magnetic, I-M, and quantum phase transitions. Finally, we show intercalation-based promising device applications in energy-efficient neuromorphic computing and offer a perspective on future research lines both fundamentally and technologically. These contents are outlined in Figure 2.[31,34,51,72,80,86-94]

## 1. Intercalation methods

A few intercalation methods have been developed to accommodate the different characteristics of both intercalant and host materials. In general, according to the intercalant state and driving force, these methods can be grouped into four categories, including liquid-phase intercalation, electrochemical intercalation, vapor-phase intercalation, and solid-phase intercalation, as shown in Figure 3 and Table 1.

### 1.1 Liquid-phase intercalation

Liquid-phase intercalation is achieved by mixing hosts and intercalants in a solvent, and intercalation takes place spontaneously without an external electric potential, as shown in Figure 3A. The adopted intercalants can be atoms, ions, or molecules since all of them are capable of existing in solvents. The intercalation level can be roughly modulated by varying the concentration of the intercalant, while the intercalation process can be accelerated by blending or heating the solvent.

Metal atoms can be intercalated into layered materials directly by a process known as zero-valent intercalation. For example, through disproportionation redox reactions or hydrazine reduction in solutions, some zero-valent metal atoms, such as Ag, Au, Cu, Co, Fe, Ni, In, and Sn, are intercalated into $Bi_2Se_3$ nanoribbons.[95,96] Furthermore, dual-element intercalation of zero-valent metals was also achieved in $Bi_2Se_3$ nanoribbons using similar methods.[97] Motter et al. proposed a general strategy to intercalate zero-valent Cu atoms into other 2D materials, including $MoO_3$, $Sb_2Te_3$, $In_2Se_3$, and GaSe, expanding the host family of zero-valent intercalation.[98] Another series of metal atoms, including semiconductor, semimetal, and heavy metal atoms, were successfully intercalated into 2D layered chalcogenides ($Bi_2Se_3$ and $NbSe_2$) using reduction with stannous chloride or decomposition of zero-valent coordination compounds in solutions.[99] In addition, the zero-valent intercalation was further extended to the microscale. Lateral p-n-metal junctions were fabricated by injecting Co and Cu atoms into different locations of the same $SnS_2$ flakes by using electron beam lithography and liquid-phase intercalation.[100]

It is possible to introduce small alkali ions, such as $Li^+$, $Na^+$, and $K^+$ ions, directly into layered materials in solutions. For example, $MoS_2$ and $ZrSe_2$ were intercalated with $Li^+$ ions by immersing them in n-butyl lithium solutions,[90,101] and $K^+$ ions were injected into MXene by dispensing $Ti_3C_2T_x$ flakes in an aqueous KOH solution.[102] In another example, Li (Na) and K metals were dissolved in liquid ammonia and then intercalated into graphene[103] and $MoS_2$[104], respectively. Moreover, alkali ions can be used to prepare ultrathin layered materials by intercalation and then exfoliation.[89]

In addition, molecules can also be introduced into layered materials using the liquid-intercalation method. For example, chiral molecules, R-α-methylbenzylamine and S-α-methylbenzylamine, can be inserted into H-$TaS_2$, T-$TaS_2$, or T-$TiS_2$ by stirring their mixture in specific conditions.[88] Other molecules, such as isopropylamine, hydrazine, urea, and dimethyl sulphoxide, have been successfully intercalated into MXenes.[105,106]

### 1.2 Electrochemical intercalation

In the process of electrochemical intercalation, an external electric potential provides the driving force for the intercalation process, which occurs in a typical electrochemical cell. The host material is designated as the cathode, while the intercalant is the ions in the electrolyte (Figure 3B). In comparison to other intercalation methods, electrochemical intercalation is reversible under positive and negative biases, and the level of intercalation can be precisely controlled by the magnitude of the bias. Therefore, this method is advantageous in those applications that necessitate delicate manipulation, in situ response, and reversible control of the intercalation. The dual-polarity external electric potential allows for the intercalation of either an anion or a cation into host materials for a variety of applications.

Anion intercalation has two main applications, which are the exfoliation of layered materials and rechargeable batteries. In the case of exfoliating layered materials, the most commonly used anion, for instance, $SO_4^{2-}$ can produce gases of $O_2$ and $SO_2$ during the intercalation process.[107] These gases largely expand interlayer gaps, thus assisting the exfoliation process followed. To date, $SO_4^{2-}$ has been employed to exfoliate a range of layered materials, such as graphite,[107,108] TMDs,[109,110] and black phosphorus.[111] Moreover, different electrolytes containing different acids have been examined to improve the exfoliation efficiency, e.g., the electrolyte made from $H_2SO_4$ shows the highest value.[108] However, due to the strong oxidizing property of $H_2SO_4$, the exfoliation products always exhibit a greater number of defects, rendering them unsuitable for use in electronic devices.[108] In the scenarios of rechargeable batteries, graphite is typically employed as the cathode electrode, which is intercalated with various anions. For example, in the case of Al-ion battery, $AlCl_4^{-1}$ and $Al_2Cl_7^{-1}$ are intercalated into graphite,[112,113] whereas in the Li-ion battery, other anions (such as $PF_6^-$, $BF_4^-$, bis(fluorosulfonyl)imide anion, and bis(trifluoromethanesulfonimide) anion) are intercalated into graphite.[114] In another application, reversible electrowetting has been achieved through the intercalation/deintercalation of $ClO4^-$ or bis(trifluoromethanesulfonyl)imide anions into/from graphite using the electrochemical method.[115]

Cation intercalation is a technique frequently employed to regulate the electronic properties of channel materials in solid-state devices. Light cations with small radii, such as the proton ($H^+$) and alkali ions, are commonly utilized because they can be reversibly intercalated/deintercalated while maintaining the lattice frameworks of the host lattices.[26] The intercalation of $Li^+$ and $H^+$ into layered materials or heterostructures has been employed to manipulate optical and electronic properties,[36,37,49,91,116] and tune the critical temperatures of superconductors[117,118] and the Curie temperatures of ferromagnetic materials.[28,119] Moreover, organic ions also have been intercalated into TMDs with the objective of creating or tailoring their superconductivity.[120-122] It is important to note that the device channel is susceptible to defect states, necessitating the electrolyte to be mild and devoid of oxidizing properties. In most cases, the electrolyte is typically in a liquid state, which facilitates the intercalation process. Nevertheless, in some instances, the utilization of a solid electrolyte is preferred for the following reasons.[2] Firstly, in low-temperature measurements, the glass transition of liquid electrolytes may deform and even damage the sample, while this drawback can be prevented in solid electrolytes.[123,124] Secondly, the fabrication of the solid electrolyte substrate is compatible with the complementary metal oxide semiconductor (CMOS) process, which suggests that electrochemical transistors on solid electrolytes may have potential applications in integrated circuits.[77] Thirdly, the selected ions can be intercalated into the host material from the bottom surface, leaving the upper surface uncovered. This enables the in-situ surface characterization of the intercalation.[54]

As with liquid phase intercalation, molecules can also be inserted into hosts by the electrochemical method. For example, organic molecules (cetyl-trimethylammonium bromide (CTAB)) have been intercalated into multilayer black phosphorus and $MoS_2$.[125,126] The incorporation of large molecules into the interlayer space results in an expansion of the interlayer gaps and the decoupling of neighboring layers. The multilayer black phosphorus undergoes a decoupling process, resulting in the formation of monolayer phosphorene and eventually, phosphorene/CTAB superlattices.[125] Similarly, $MoS_2$ thin films are decoupled to become bulk monolayer $MoS_2$ and show heavy electron doping.[126] After the introduction of a poly(vinylpyrrolidone)–bromine complex to de-dope the bulk monolayer $MoS_2$, the thin films exhibited excellent optical and photoelectric characteristics.[126]

### 1.3 Vapor-phase intercalation

In the process of vapor-phase intercalation, foreign species are introduced into hosts via vapor form, and temperature is employed to control the vapor pressure of the intercalant and the vapor-transport process. Normally, this intercalation is divided into two categories: the isothermal technique and the two-zone technique.[3] In the former technique, the host and intercalant are sealed in the same vessel, maintaining the same temperature. In the latter technique, the host and intercalant are sealed in the same vessel but in two different temperature zones, as shown in Figure 3C.[3] Vapor-phase intercalation is a universal technique that can be applied to a wide range of hosts and intercalants, with the potential to enable large-scale production. Nevertheless, the intercalation process utilizing this method is not reversible, which presents a challenge in certain application scenarios, such as sensors and rechargeable batteries.

In certain instances, maintaining a temperature differential between the host and the intercalant is not necessary for the intercalation process to proceed. Consequently, an isothermal technique with a simple apparatus is employed. With this technique, $AsF_5$ was used to intercalate into graphite and then formed three stages of compounds, dependent on the reaction time.[127] Moreover, the $AsF_5$ system was the first to demonstrate spontaneous staging under isothermal conditions.[127] Additionally, alkalis were employed to intercalate into few-layer graphene using the isothermal technique. Both K (heated at 210 °C) and Rb (heated at 160 °C) were intercalated between graphene layers and absorbed on the graphene surfaces, resulting in the activation of new intrinsic Raman modes and the transfer of electrons to graphene.[128]

The intercalation of graphite and graphene with halogen vapors was achieved using the two-zone technique, and the resulting structural evolution was monitored through the use of Raman spectra.[129] After the intercalation of $Br_2$, a commensurate-incommensurate phase transition and a continuous melting transition were observed in stage-3 graphite-bromine compounds and intercalated bromine layers, respectively.[129] Upon exposure to $I_2$ and $Br_2$ vapors, 1- to 4-layer graphene exhibited different absorption and intercalation behaviors.[130] While $Br_2$ was absorbed on 1- and 2-layer graphene and intercalated into 3- and 4-layer graphene, $I_2$ was only absorbed on 1- to 4-layer graphene without intercalation.[130] The enhancement of Raman spectra observed in graphene exposed to $I_2$ and $Br_2$ vapors can be attributed to intramolecular electronic resonance, which arises from the adsorption of iodine and intercalation of bromine.[131] In another example, $FeCl_3$ was also intercalated into few-layer graphene employing the two-zone technique.[132,133] The graphene layers were found to be decoupled and highly doped, exhibiting high conductance and high transparency comparable to that of the widely used transparent conductor indium tin oxide.[132]

### 1.4 Solid-phase intercalation

Solid-phase intercalation is a technique whereby a mixture of solid hosts and solid intercalants is heated at the same temperature zone, as shown in Figure 3D. This method is rarely employed due to the uncontrollable nature of intercalation between two solids, resulting in a slow intercalation process and the formation of inhomogeneous products.

In general, to accelerate intercalation reactions and achieve homogeneous intercalation productions, the heating temperature of solid-state intercalation should be high. For example, at 800 °C, the intercalated $Cu_{0.65}NbS_2$ compound was obtained by heating the mixture of Cu and $NbS_2$ in an evacuated quartz tube for several days.[134] Nevertheless, the high-temperature intercalation process inevitably introduces defects into the host materials and compromises the quality of the resulting intercalation compounds. One potential solution to this issue is to grind the granular intercalant into powder, which would facilitate diffusion at a relatively low temperature. For example, a blend of Cu powder and $TiS_2$ polycrystals was pressed to reduce the diffusion length, thereby facilitating intercalation.[135] This resulted in the formation of layered compounds based on the $TiS_2$ structure at room temperature. In a similar manner, Cd was intercalated into $Ti_2PTe_2$ at 80 °C, and Cu was intercalated into $Zr_2PTe_2$ at 70 °C.[136]

### 1.5 Characterization techniques for intercalation

Intercalation is known to result in crystal expansion and structural deformation, which can be characterized by a range of techniques (Figure 4), including TEM, STM, AFM, XRD, XPS, and Raman spectroscopy. Other techniques, such as resistance measurement, photoluminescence (PL) spectroscopy, transmission spectroscopy, second harmonic generation (SHG), and secondary ion mass spectrometry, can be employed to examine specific phase transitions. The advantages and disadvantages of these techniques are systematically summarized in Table 2.

TEM imaging is based on a high-energy electron beam and can directly image the ultrafine structure of crystals in real space. This provides convincing evidence for phase transitions, as well as detailed crystal parameters during phase transitions. Two operational modes are available for TEM: conventional TEM mode and scanning TEM (STEM) mode. The former with a relatively low resolution, is employed to image bulk structures, such as crystal defects. In contrast, the latter, with a high resolution, is utilized to image vdWs structures in different phases on the atomic scale. Conventional TEM allows for the characterization of microstructures before and after intercalation, such as the intercalation homogeneity and the change of layer spacing. Zhao et al. have demonstrated the ability to trigger a quantum Griffiths singularity in $SnSe_2$ by intercalation of tetraoctylammonium cations ($TOA^+$), with a concomitant increase in the layer spacing from 6.1 Å to 15.2 Å observed using conventional TEM (Figures 4A and 4B).[87] Moreover, STEM is frequently combined with the HAADF (high-angle annular dark-field) detector to enhance imaging resolution. For example, Li et al. have intercalated $Co(Cp)_2$ molecules into $SnSe_2$ and obtained the cross-sectional image (Figure 4C) and the basal-plane image (Figure 4D) of the $SnSe_2$−$Co(Cp)_2$ superlattice at atomic resolution using HAADF-STEM.[50] In addition, the technique of selected-area electron diffraction (SAED) can be employed in both conventional TEM and STEM to identify monocrystalline, polycrystalline, and amorphous materials. The lattice parameters can be extracted from the SAED patterns before and after intercalation.

STM can also acquire atomic-resolution images of the topography based on the quantum tunneling effect. However, due to the differing operational principles, STM is only capable of capturing the surface structure, whereas TEM is able to image the internal structure. Therefore, STM is suitable for characterizing the intercalation-induced structural variation on the top layer. For instance, Pan et al. employed STM to monitor the phase transition of 1T-$NiTe_2$ under self-intercalation of Ni.[137] Figures 4E and 4F illustrate the STM topography of the pristine 1T-$NiTe_2$ and the self-intercalated $Ni_3Te_4$, respectively.[137] These images demonstrate the direct observation of the phase transition at the atomic level. Furthermore, the high sensitivity of STM enables the detection of even small variations in the superconducting or CDW energy gap after phase transitions induced by intercalation.[138,139] To ascertain whether the energy gap is the result of the superconducting or the CDW phase transition, a magnetic field of sufficient strength to destroy the superconducting state is typically applied during the measurement of the scanning tunneling spectroscopy spectra.[139]

AFM is another technique for characterizing the topography of intercalated materials, albeit with a lower resolution than the aforementioned techniques. Normally, it is employed to measure the crystal thicknesses, and the extracted expansion ratio during intercalation provides an indication of the intercalation level. For example, Qian et al. employed AFM to validate the intercalation of chiral molecules into H-$TaS_2$ by comparing the crystal thicknesses before (Figure 4G) and after (Figure 4H) intercalation, which demonstrated an expansion ratio of approximately 100%.[88] It is worth noting that the resolution of an AFM is largely dependent on the radius of the AFM probe, and the use of a probe with a large radius may introduce artifacts into the data. In addition, other operational modes of AFM, such as piezoresponse force microscopy (PFM) and magnetic force microscopy, have been employed to characterize the ferroelectric phase transition[54] and the ferromagnetic phase transition[30] induced by intercalation, respectively.

XRD characterization is based on Bragg's Law and X-ray whose wavelength (0.1 nm ~ 10 nm) is comparable with the lattice spacing (~ 0.1 nm). The technique provides information regarding materials, such as lattice parameters, grain sizes, dislocation densities, and film thicknesses. Since the symmetry, the interlayer spacing, the phase ratio, and the defect quantities, etc. can be altered during intercalation, XRD is thus a favorable technique to monitor the phase transition. For example, the distinct XRD patterns indicate the varying lattice spacings of pristine $MoS_2$ and expanded $MoS_2$ intercalated with diverse molecules, as shown in Figure 4I.[125] Additionally, XRD can demonstrate the evolution of the intercalated phase in relation to the intercalation time, temperature, or the gating voltage during intercalation by continuously recording the shift of the peak positions.[29,34]

XPS is another non-destructive technique based on X-ray irradiation, yet is only applicable to surface analysis with a depth between 1 nm and 10 nm. The collection of the ejected electrons excited by an X-ray beam allows the distribution and the chemical states of surface elements to be obtained. Generally speaking, intercalation occurs primarily at the top surface, where the structural changes are most pronounced, thus facilitating the XPS characterization. For example, XPS has been utilized to differentiate between distinct phase compositions in $MoS_2$ after the intercalation of two distinct types of molecules (Figure 4J).[125] Despite the advantages of XPS in material science, there are several limitations to this technique, including its relatively low spatial resolution and inability to detect hydrogen and helium elements. Thus, this technique is unsuitable for demonstrating the common proton intercalation.

Raman spectroscopy is a high-resolution (in both space and spectrum) and non-destructive optical technique, which has been widely applied in studies of phase transitions induced by intercalation. The Raman spectrum is generated by illuminating a laser on samples and then collecting emitted photons that inelastically interact with molecules of the sample at different frequencies. Each vibration mode of the crystal is associated with a specific peak in the Raman spectrum. Since intercalation leads to carrier doping and the formation of new chemical bonds,[2] which directly affect the crystal vibrational modes, it can thus be reflected in the shift of the Raman peaks. For example, the Raman peaks of pristine $SnSe_2$ undergo a redshift after intercalation of tetraoctylammonium cations.[87] In addition, Raman spectra can also indicate phase transitions by the presence of new characteristic peaks resulting from the introduction of new vibration modes.[80,116] For example, as shown in Figure 4K, Raman peaks at 384 $cm^{-1}$ and 404 $cm^{-1}$ correspond to the in-plane mode ($E_{2g}^1$) and the out-of-plane mode ($A_{1g}$) of pristine 2H-$MoS_2$, respectively.[80] After intercalation of $Li^+$ ions, additional peaks appear at 200, 225 and 355 $cm^{-1}$, indicating the emergence of 1T'-$MoS_2$.[80]

A variety of additional techniques may be utilized to characterize intercalations, such as resistivity measurement, PL spectroscopy, transmission spectroscopy, SHG, and secondary ion mass spectrometry. In particular, resistance measurement has been extensively utilized to demonstrate I-M phase transitions,[23] magnetic phase transitions (Figure 4L),[86] CDW phase transitions,[71] and superconducting phase transitions (Figure 4M);[121] PL spectroscopy has been employed to distinguish different intercalation stages (Figure 4N)[125] and different phases induced by intercalation.[140] Transmission spectroscopy has been used to demonstrate the electrochromic effect induced by intercalation (Figure 4O);[49] SHG has been employed to investigate the alteration of the crystal symmetry induced by intercalation,[34] while secondary ion mass spectrometry has been utilized to quantify the depth of intercalation.[24]

## 2. Intercalation-based phase transitions

The introduction of foreign species into host materials is accompanied by lattice expansion, charge transfer, or chemical bonding.[2] Lattice expansion can correspondingly modulate the strength of chemical bonds and the band structures of host materials, which may result in the host relaxing to a more stable phase. The transfer of charge between intercalants and hosts can induce charge doping, resulting in alterations to the conductivity, the conduction type (i.e., P or N type), and even a phase transition in the host. The formation of new chemical bonds can alter the crystal structure and symmetry of a host material, as well as the energy band structure. This process can result in the evolution of the material into a new phase. Building upon previous research,[2,141] the phase transition discussed in this review consists of ferroic, electronic, and structural phase transformation. With respect to order parameters (e.g., electric polarization, magnetization, and strain), the ferroic phase transition should encompass ferroelectric, ferromagnetic, and ferroelastic counterparts.[141] Nevertheless, the ferroelastic phase transition induced by intercalation is a rare occurrence and will be introduced in conjunction with the ferroelectric phase transition. The electronic phase transition primarily involves dramatic changes in conductivity, including I-M phase transitions and quantum phase transitions (i.e., the superconducting and CDW phase transitions). In experiments, the quantum phase transition can only be observed in strongly correlated systems, where the electron-electron interactions cannot be ignored. It is noted that structural phase transitions always occur along with other types of phase transitions. Therefore, they will not be introduced separately.

### 2.1 Ferroelectric phase transitions

Ferroelectric materials have been studied extensively due to their intriguing physical properties and significant industrial applications,[142,143] e.g., data storage. The intercalation of ferroelectric materials could form new chemical bonds and change the symmetry of the initial lattice, leading to the emergence of new ferroelectric phases. Moreover, the charge doping through intercalation can also influence the long-range interactions between electric dipoles,[144] and eventually affect the ferroelectric phase transition.

The majority of traditional ferroelectric materials exhibit poor conductivity, so they normally act as the dielectric layer in a ferroelectric field-effect transistor with channel carriers depleted or accumulated by the polarization state of the ferroelectric layer.[145] Si et al. developed a ferroelectric semiconductor field-effect transistor with semiconducting ferroelectric α-$In_2Se_3$ as the conduction channel, which can eliminate the charge trap and the leakage current issues commonly observed in the conventional ferroelectric field-effect transistor.[146] However, only a few resistance states can be acquired in these transistors due to the limited domain configurations, which restrict the increase of memory capacity. Our group has developed a novel approach to induce ferroelectric phase transitions and significantly increase the resistance states in an α-$In_2Se_3$ electrochemical transistor.[54] An α-$In_2Se_3$ thin flake was transferred onto a solid electrolyte of porous $SiO_2$ above a Pt electrode and subsequently patterned into an electrochemical transistor (Figure 5A). The porous $SiO_2$ layer serves as a standard proton reservoir, capable of injecting protons into the α-$In_2Se_3$ channel under a positive gate voltage (α-$In_2Se_3$+x$H^+$+x$e^-$→α-$In_2Se_3H_x$), or extracting protons from the α-$In_2Se_3$ channel under a negative gate voltage (α-$In_2Se_3H_x$→α-$In_2Se_3$+x$H^+$+x$e^-$). When the top surface of the porous $SiO_2$ layer is smooth, efficient injection or extraction of protons can be achieved even with a small gate voltage, corresponding to a rapid change of the channel conduction in the transfer curve (Figure 5B).[54] The emergence of $SiO_2$ particles at the α-$In_2Se_3$/$SiO_2$ interface has been observed to result in a significant reduction in the efficiency of proton injection/extraction, as evidenced by the gentle slope of the transfer curve (Figure 5C).[54] Moreover, the α-$In_2Se_3$ channel exhibits distinct proton concentrations and different ferroelectric phases under varying gate voltages, which manifests disparate electromechanical responses. Thus, these ferroelectric phases can be detected by the various brightnesses of the PFM amplitude mapping illustrated in Figure 5C.[54] It is worth noting that these different ferroelectric phases are associated with different resistance states, holding immense potential for increasing memory capacity.

Furthermore, intercalation can also trigger a transformation between different ferroic phases. For example, Wu et al. have synthesized a ferroelectric hybrid perovskite $(TMA)_3Sb_2Cl_9$ (TSC) (TMA represents trimethylammonium), and it can be transformed into a ferroelastic material $(TMA)_4$-Fe(iii)$Cl_4$-$Sb_2Cl_9$ (TSFC) by intercalation of $FeCl_4^-$, which alters the symmetry of the parent compound TSC (Figure 5D).[53] The ferroelectric hysteresis loops of TSC measured at room temperature exhibit a polarization value of approximately 1.76 μC/$cm^2$. Additionally, the PFM amplitude and phase signals display spontaneous ferroelectric strip domains, indicating unambiguous ferroelectric behavior.[53] While the domain walls of TSFC have been observed using polarized light optical microscopy (Figure 5E), and they respond to external stresses instead of electric fields, manifesting ferroelastic behavior These ferroelastic domain walls expand upon upward bending (Figure 5F), shrink upon relief, and vanish upon downward bending, displaying remarkable ferroelastic behavior.[53] This new avenue could be used to explore the multiferroic properties of layered perovskite and promisingly realize multifunctional devices based on layered perovskite.

Surprisingly, ferroelectricity can be created in a non-ferroic material by intercalation. Lipatov et al. have transformed semiconducting 2H-$MoS_2$ into metallic1T''-$MoS_2$ by intercalation of $Li^+$ ions, and 1T''-$MoS_2$ displays ferroelectricity.[52] The calculated electronic charge density for polar 1T''-$MoS_2$ and a reference centrosymmetric structure, along with the charge density difference between them, can be found in Figure 5G. This figure illustrates the formation of a dipole by the "polar" S atoms, elucidating the origin of the polarity in 1T''-$MoS_2$. Due to the high conductivity of 1T''-$MoS_2$, its polarization cannot be switched directly by an external electric field. Instead, the polarization can be reversed by strain gradients via the flexoelectric effect.[52] The strain gradient-induced domain switching can be demonstrated by the enhanced PFM amplitude in Figure 5H and the 180°-reversed PFM phase in Figure 5I. In addition, a control experiment was performed on non-ferroelectric 2H-$MoS_2$ using the same mechanical switching method, and no change was observed in either the PFM amplitude or the PFM phase signal.[52] Although a ferroelectric metal has been discovered, more non-intrinsic ferroelectric metals with unprecedented physics could be created using this novel method. In addition, further research is needed to realize electric field switchable polarization for non-intrinsic ferroelectric materials created by intercalation.

### 2.2 Magnetic phase transitions

Magnetic phase transitions can also be triggered by intercalation, since intercalation-induced charge doping can modify the spin-orbit coupling, the interlayer magnetic coupling, or the electronic band structure, while the new chemical bonds induced by intercalation can induce unpaired electrons and then increase the magnetic moments.[2,11] In addition, intercalation can be used to enhance the Curie temperature of 2D ferromagnets,[28] which benefits the application of magnetic devices at room temperature.

The paramagnetic $Ho^{3+}$ ion exhibits large magnetic susceptibility due to the four unpaired electrons in the 4f shell, and it has been intercalated into various layered 2D materials to induce phase transitions from the non-magnetic phase to the paramagnetic phase.[32] Taking graphene oxide (GO) as an example, after immersing GO flakes in a $Ho(NO_3)_3$ solution, $Ho^{3+}$ ions can intercalate into GO to form $Ho^{3+}$-GO complexes by coordinating with the carboxylate groups, or other oxygen-containing groups and aromatic rings on the surfaces of GO, as shown in Figure 6A.[32] The magnetic response of these materials after intercalation of $Ho^{3+}$ ions can be modulated by the immersion time in the $Ho(NO_3)_3$ solution. For example, the volume susceptibilities of a $Ho^{3+}$-GO film are increased with the immersion time: from $0.63 \times 10^{-4}$(the pristine GO) to $3.53 \times 10^{-4}$ (0.5 h), $3.73 \times 10^{-4}$ (1 h), $4.32 \times 10^{-4}$ (2 h), and $4.86 \times 10^{-4}$ (4 h), indicating the enhancement of the $Ho^{3+}$-GO magnetism.[32] Interestingly, the magnetism induced by the $Ho^{3+}$-ion intercalation is stable even in deionized water and base conditions (Figure 6B), but unstable in strong acidic conditions (Figure 6C). This is because protons in acid can replace $Ho^{3+}$ ions to coordinate with carboxylate groups, leading to the deintercalation of $Ho^{3+}$ ions. In particular, the $Na^+$-EDTA solution can also remove the magnetism of $Ho^{3+}$-GO in a mild condition (PH $\approx$ 7) (Figure 6B).[32] Such a magnetization method is simple, general, and scalable, thus promising for practical magnetic devices. Moreover, it is meaningful to explore whether diamagnets (e.g., graphite) and ferromagnets (e.g., $Fe_3GaTe_2$) can be transformed into paramagnets by intercalation of $Ho^{3+}$ ions in future studies.

Wang's group has developed an effective method for modulating the magnetic properties of layered ferromagnetic materials or inducing magnetic phase transitions through the intercalation of protons.[33,35,147,148] For example, Tan et al. have injected protons into a quasi-2D magnet $Cr_{1.2}Te_2$, whose Curie temperature is as high as 320 K.[33] The evolution of the anomalous Hall effect (AHE) loop with the gate voltage at 200 K can be seen in Figure 6D, where the coercivity varies with the gate voltage in a nonmonotonic manner and vanishes at the gate voltage of -14 V.[33] Furthermore, the amount of injected protons increases with an increase in the gate voltage, with each injected proton accompanied by an electron to maintain the global charge neutrality. Consequently, the hole density of $Cr_{1.2}Te_2$ thus decreases with an increase in the gate voltage, reaching a reduction of $3.4 \times 10^{21}$ $cm^{-3}$ at a gate voltage of -14 V.[33] According to the density functional theory calculations, the vanishment of the AHE loop and the coercivity can be attributed to the phase transition of $Cr_{1.2}Te_2$ from the ferromagnetic phase to the antiferromagnetic phase by electron doping.[33] In another example, Tan et al. also employed proton intercalation to trigger the phase transition of $Fe_5GeTe_2$ from the ferromagnetic phase to the antiferromagnetic phase, at an electron doping concentration higher than $10^{21}$ $cm^{-3}$.[35] In these examples, the magnetic phase transitions are triggered in metals by the variation in carrier density. This implies that intercalation-induced doping offers significant advantages over other doping strategies (e.g., electrostatic gating), especially in thick materials with high intrinsic carrier densities.

In other examples, protons are employed to initiate the magnetic phase transition as well. Wang et al. employed ionic liquid gating to modulate the electronic and magnetic properties of inverse spinel $NiCo_2O_4$ through proton intercalation.[29] It was observed that elevated temperature contributes to the phase transition of gated $NiCo_2O_4$, evidenced by the XRD spectra at different temperatures (Figure 6E).[29] Subsequently, X-ray magnetic circular dichroism measurements were conducted on the Ni (Figure 6F) and Co (Figure 6G) *L*-edges, which revealed that the spin polarizations for both Ni and Co ions in the pristine $NiCo_2O_4$ have vanished in the protonated $NiCo_2O_4$, implying an antiferromagnetic phase.[29] In addition, the relationship between the resistance of $NiCo_2O_4$ and temperature was investigated by measuring the resistance-temperature curves. This revealed that $NiCo_2O_4$ has undergone a transformation from a ferromagnetic metal to an antiferromagnetic insulator.[29] Li et al. have verified proton-intercalation-induced phase transition in $SrRuO_3$ from an exotic ferromagnetic phase to a paramagnetic phase, and such a phase transition is dynamically reversible when the gate voltage is switched between 0 V and 3.5 V.[34] Furthermore, Shen et al. have observed a crossover from a non-Fermi-liquid to a Fermi-liquid state in $CaRuO_3$ induced by ionic-liquid-mediated proton intercalation, accompanied by a magnetic phase transition from a paramagnetic phase to an exotic ferromagnetic phase.[86] Proton intercalation is a preferred method for studying magnetic phase transitions since the proton is the smallest and lightest ion with high diffusivity. This allows for the intercalation of protons into magnetic metals, enabling doping and rapid phase transitions.

### 2.3 I-M phase transitions

Intercalation can lead to I-M phase transitions through charge doping and structural phase transition. This is due to the fact that intercalation-induced high-level charge doping can result in the shift of the Fermi level to the conduction band, thereby converting a semiconductor to a degenerate semiconductor with metallicity,[23] accompanied by intercalation-induced structural deformation. Alternatively, intercalation can directly transform a semiconducting phase into a metallic phase via structural phase transition,[140] accompanied by both charge transfer and strain effects.

It has been demonstrated that various intercalants can facilitate I-M electronic phase transitions. For example, Wang et al. have introduced organic cations into a ferromagnetic semiconductor, $Cr_2Ge_2Te_6$, thereby transforming the compound into a ferromagnetic metal with a higher Curie temperature.[59] The organic ions of tetrabutyl ammonium ($TBA^+$) were intercalated into $Cr_2Ge_2Te_6$ using electrochemical intercalation, forming the $(TBA)Cr_2Ge_2Te_6$ hybrid superlattice, as depicted in Figure 7A. Given that the size of $TBA^+$ is as large as 10 Å, $(TBA)Cr_2Ge_2Te_6$ was found to be considerably expanded compared with the pristine $Cr_2Ge_2Te_6$, whose interlayer distance is only 6.8 Å. It is noteworthy that the large molecules not only act as dopants for the host material but also serve to reduce the interaction between adjacent layers.[59] Surprisingly, the Curie temperature of $(TBA)Cr_2Ge_2Te_6$ reaches 208 K, much higher than that of the pristine $Cr_2Ge_2Te_6$ (67 K) (Figure 7B). This can be ascribed to a change in the magnetic coupling from a superexchange interaction to a double-exchange interaction.[59] The intercalation process has resulted in alterations to the magnetic and electronic properties of the material. As shown in the resistance-temperature curves in Figure 7C, $Cr_2Ge_2Te_6$ has undergone a semiconductor-metal transition after $TBA^+$-intercalation. Interestingly, after annealing at 150°C under vacuum for two hours to extract the organic $TBA^+$ cations, both the magnetic and electronic properties of $Cr_2Ge_2Te_6$ exhibit a near-complete reversal to their initial states.[59] Similarly, Meng et al. have introduced $TBA^+$-intercalation into $VSe_2$, which also leads to a metal-insulator transition, as well as the enhancement of the transition temperature of the CDW phase.[149] Therefore, the intercalation of large molecules provides a novel approach to decouple layered materials and modify their magnetic or other properties simultaneously besides doping, which indicates a distinct strategy for inducting phase transitions.

In other examples, protons have been employed to initiate the semiconductor-metal phase transition as well. For example, Jo et al. have introduced proton intercalation into $VO_2$ to induce a two-step phase transition from an insulating state to a metallic state, and then back to an insulating state.[26] The experimental setup for the realization of the phase transition is shown in Figure 7D, which depicts an electrochemical transistor comprising a $VO_2$ channel and an electrolyte of porous silica. The porous silica serves as a proton reservoir, capable of absorbing water vapor from the air and electrolyzing it into $H^+$ and $OH^-$ ions upon application of a gate voltage exceeding 1.23 V, the standard electrode potential for water electrolysis.[26] Moreover, the protons can be intercalated into the $VO_2$ channel under a positive gate voltage, while they can be extracted from the $VO_2$ channel under a negative gate voltage. When the gate voltage is swept in a range between -3 V and 3 V, the transfer curve exhibits three states: the initial insulating $VO_2$ (i-$VO_2$) state at 0 V, the metallic $H_xVO_2$ (m-$H_xVO_2$) state at 2 V, and another insulating $VO_2$ (i-$H_yVO_2$) state at 3 V (Figure 7E), corresponding to an insulator-metal-insulator transition.[26] Such reversible phase transitions can also be demonstrated by the evolution of the channel conductance with time under either the large positive gate voltage or the large negative gate voltage, as shown in Figure 7F.[26] As the intercalation of protons here is facilitated by the absorption of water from the surrounding atmosphere, and thus subject to environmental influences, further studies are required to optimize the device structure in order to mitigate external influences, e.g., by ensuring that sufficient protons are stored within the dielectric layers of packaged devices.

In another example, Wang et al. used ionic liquid gating to investigate the insulator-to-metal transition in $WO_3$.[60] The temperature-dependent resistance under different gate voltages is shown in Figure 7G. The pristine $WO_3$ displays typical insulating behavior, with the resistance of $WO_3$ decreasing with increasing the gate voltage. The I-M transition occurs at approximately 1.5 V, where the trend of increasing resistance with decreasing temperature is suppressed. As the gate voltage is increased to a value of 3.5 V or below, the resistance is significantly reduced. Nevertheless, a further increase in the gate voltage leads to an enhancement of the $WO_3$ resistance, despite the continued increase in carrier density.[60] To elucidate this anomaly, in-situ XRD measurements were conducted during ionic liquid gating. The 2θ−ω scan around the $WO_3$ (400) peak indicates a structural phase transition at approximately 3.5 V (Figure 7H), which is consistent with the gate voltage at which the observed abnormality occurs in Figure 7G.[60] Moreover, the phase transition can be attributed to the proton intercalation induced by the electrolysis of the water present in the as-received ionic liquid. This conclusion is supported by the isotope labeling with heavy water ($D_2O$), as illustrated in Figure 7I.[60] The aforementioned examples illustrate that proton intercalation represents an effective approach to modifying the electronic properties of d-orbital transitional metal oxides (such as $VO_2$ and $WO_3$), as the proton is capable of bonding with oxygen and modulating the d-orbital occupancy, thereby facilitating a transition between electron-localized states and electron-itinerant states,[150] i.e., the insulating phase and the metallic phase.

## 2.4 Quantum phase transitions

### 2.4.1 Superconducting phase transitions

Superconductivity is sensitive to the carrier density and generally emerges at a carrier density exceeding $10^{14}$ $cm^{-2}$.[2] Doping-based superconductivity has been achieved through electrostatic doping from ionic liquid gates; however, the superconducting phase only manifests in the vicinity of the surface, frequently resulting in the segregation of the normal phase and the superconducting phase.[151] In contrast, intercalation can facilitate the creation of a higher carrier density and a more uniform distribution of the superconducting phase across the sample. This is due to the fact that intercalants can be introduced to each interlayer, thereby doping the adjacent layers.

The superconducting phase transition in $MoS_2$ was first observed in 1971 by intercalation of Na and K.[65] In the same year, the intercalation of organic molecules into $TaS_2$ (an intrinsic superconductor) was found to enhance the superconducting transition temperature.[152] Whether the intercalation of organic molecules can transform $MoS_2$ into a superconductor has become a long-term unsolved issue. In a recent study, Pereira et al. employed the electrochemical method to intercalate two types of organic molecules (cetyltrimethylammonium ($CTA^+$) and tetraethylammonium ($TEA^+$)) into $MoS_2$, respectively (Figure 8A).[120] $(CTA)_xMoS_2$ shows a resistance drop at around 2.8 K, yet it does not reach a zero-resistance state even when the temperature is reduced to 1.8 K (Figure 8B). This can be ascribed to the non-homogeneous doping of $(CTA)_xMoS_2$, which had been validated by the experiments conducted using both the scattering-type scanning near-field optical microscopy and the nanoscale Fourier-transform infrared nanospectroscopy.[120] In contrast, $(TEA)_yMoS_2$ displays a complete superconducting phase transition with an onset temperature of approximately 3.9 K, as shown in Figure 8C. Although the transition temperature of $(TEA)_yMoS_2$ is lower than that of $MoS_2$ intercalated with alkali metals,[65] its superconducting transition can persist for as long as six months in air.[120] This stability is a notable improvement over that of K-intercalated $MoS_2$, suggesting a promising avenue for stabilizing the intercalated superconducting phase.

β-ZrNCl is another typical layered semiconductor that has been transformed into a superconductor, $Li_xZrNCl$ (Figure 8D),[118] by Li-ion intercalation. Yamanaka first demonstrated that the semiconductor β-ZrNCl could be transformed into a type-II superconductor $Li_xZrNCl$, through intercalation of $Li^+$ ions via immersion in a n-butyllithium (Bu-Li) solution in hexane.[153] The critical temperature of the intercalation product $Li_xZrNCl$ was found to reach 12.5 K ($x = 0.16$). Given the difficulty in synthesizing a single-phase sample of $Li_xZrNCl$ with a low intercalation level, the electronic properties of $Li_xZrNCl$ ($x < 0.16$) had not been explored until 2006. In 2006, Taguchi et al. succeeded in overcoming the previous limitations and synthesized the $Li_xZrNCl$ samples at a broader range of intercalation, with x values spanning from 0.05 to 0.31.[154] The sample was observed to become superconducting at low temperatures when the Li content (x) exceeded 0.05. Later, Nakagawa et al. achieved a lower intercalation level of $Li_xZrNCl$ through electrochemical intercalation, with x spanning two orders of magnitude from 0.0038 to 0.28.[118] The highest critical temperature of 19 K was identified at x = 0.011 (Figure 8E). Moreover, these intercalation studies in β-ZrNCl have demonstrated that electrochemical intercalation is the most accurate way to obtain the designated intercalation level across a broad range.

Graphite has been transformed into a superconductor by intercalation of alkali metals nearly half a century ago, although with ultralow transition temperatures.[64] In 2005, Weller et al. intercalated graphite with Yb and Ca, respectively, resulting in the formation of $C_6Yb$ and $C_6Ca$.[155] The introduction of these elements led to a significant enhancement in the superconducting transition temperatures, as illustrated in Figures 8F and 8G. It is noteworthy that the critical temperature of 11.5 K in $C_6Ca$ represents a record-high value for graphite in ambient pressure to date.[155] Similarly, few-layer graphene also has been transformed into a superconductor by intercalation of Ca atoms. For example, Ichinokura et al. have intercalated Ca atoms into bilayer graphene grown on SiC at an ultrahigh vacuum.[156] The first step involved depositing Li atoms onto bilayer graphene, forming Li-intercalated bilayer graphene ($C_6LiC_6$). Subsequently, Ca atoms were deposited to replace the intercalated Li atoms, resulting in Ca-intercalated bilayer graphene ($C_6CaC_6$). Upon repeating this cycle several times, the entirety of the sample will be converted into $C_6CaC_6$. Subsequently, the resistances of the samples were measured in situ at a high vacuum using the four-point probe method. Surprisingly, $C_6LiC_6$ only shows a slight indication of weak localization, whereas $C_6CaC_6$ displays a superconducting transition at approximately 4 K ($T_c^{onset}$) and zero resistance at around 2 K ($T_c^{zero}$).[156] Currently, the realization of superconductivity in monolayer graphene represents one of the most challenging subjects in the field of 2D materials. Theoretically, monolayer graphene is capable of holding an unconventional superconducting state when the Fermi level is close to, and even beyond, the Van Hove singularity.[157,158] Experimentally, extremely high carrier densities have been achieved in monolayer graphene via intercalation.[159] It is therefore promising to trigger the superconducting phase transition in monolayer graphene via an optimized intercalation method in the future.

In another example, Li et al. intercalated $SnSe_2$ with $Co(Cp)_2$ molecules by immersing the sample in an acetonitrile solution of $Co(Cp)_2$, thus artificially synthesizing a $SnSe_2$−$Co(Cp)_2$ superlattice.[50] Surprisingly, this superlattice possesses both superconducting and ferromagnetic characteristics. The resistance of $SnSe_2$−$Co(Cp)_2$ exhibits a decrease with decreasing temperature, reaching a sudden drop to become zero at approximately 5 K (Figure 8H), which is characteristic of a superconductor. At a temperature of 10 K above the critical temperature of the superconducting phase transition, the "S"-shape magnetization curve of $SnSe_2$−$Co(Cp)_2$ shows a magnetic hysteresis loop, which is indicative of ferromagnetism.[50] Upon the transition of $SnSe_2$−$Co(Cp)_2$ to the superconducting state at 2 K, the magnetization curve exhibits diamagnetic characteristics at low fields and ferromagnetic features at high fields, indicating the coexistence of superconductivity and ferromagnetism (Figure 8I).[50] The superconducting phase is attributed to the charge transfer from $Co(Cp)_2$ to $SnSe_2$, while the ferromagnetic phase can be ascribed to the high-spin state of the confined $Co(Cp)_2$ molecules in $SnSe_2$ interlayer gaps.[50] This study offers a unique opportunity to study superconducting spintronics based on layered materials. Further insight into the coexistence phase could be gained through theoretical studies on both phase transitions, such as the quantity of the transferred charges from each $Co(Cp)_2$ molecule and the magnetic moment of each $Co(Cp)_2$ molecule.

### 2.4.2 CDW phase transitions

The CDW phase typically emerges in low-dimensional materials and interacts with the superconducting phase. The phenomenon was first observed in 1T-$TaSe_2$ and 1T-$TaS_2$ in 1974,[160,161] and intercalation-based CDW phase transitions in 1T-$TaS_2$ were studied afterward.[68] The CDW phase transition can be demonstrated by the sudden jump in the temperature-dependent resistance and can be observed directly by TEM or STM. Since CDW is a modulation of conduction electrons accompanied by periodic lattice distortion,[162] intercalation-induced charge doping and lattice deformation can thus directly vary the periodicity of the CDW or even erase the CDW by disturbing the phonon dispersions.[163,164]

Yu et al. have intercalated $Li^+$ ions into 1T-$TaS_2$ to induce various CDW phase transitions via the electrochemical method.[71] These transitions include commensurate, nearly-commensurate, and incommensurate CDW phases. They first demonstrated that both the commensurate-to-nearly-commensurate and nearly-commensurate-to-incommensurate CDW phase transitions are markedly dependent on the thickness of the 1T-$TaS_2$ layer.

Specifically, they observed that the phase transition temperatures decreased with the reduction of the thickness, and that the phase transitions even vanished when the thickness was reduced to approximately 10 nm and 3 nm, respectively.[71] Subsequently, they employed electrochemical intercalation to modulate the phase transition, with the experimental setup depicted in Figure 8J. As the gate voltage is increased to specific values for 1T-$TaS_2$ with varying thicknesses, the resistances exhibit a sudden decline to approximately half of their initial values (Figure 8K). This phenomenon can be attributed to the nearly-commensurate-to-incommensurate CDW phase transition. Figure 8L illustrates the temperature-dependent resistance of a 14 nm-thick 1T-$TaS_2$ sample, with resistance jumps indicating the CDW phase transitions. When the gate voltage is higher than 2.6 V, the resistance is significantly reduced, which is indicative of the melting of the Mott phase. Further increases in the gate voltage result in a reduction in the temperature of the nearly-commensurate-to-commensurate CDW phase transition, and even the complete removal of the phase transition. Meanwhile, 1T-$TaS_2$ displays metallic conduction and undergoes a transition to a superconductor at 2 K. However, a higher gate voltage of $\geqslant$3.3 V reverts 1T-$TaS_2$ to the metallic phase.[71] In another example, Bhoi et al. have used Pd intercalation to initiate a CDW phase transition in 2H-$TaSe_2$ among the incommensurate CDW phase, commensurate CDW phase, and superconducting phase, which is evidenced by the temperature-dependent resistance.[165] It is noteworthy that either pure pressure[166] or pure electrostatic doping[167] can result in CDW phase transitions, whereas intercalation can induce both the strain and the doping effects, thus offering a more efficient approach to creating CDW phase transitions.

It has been demonstrated that $Ca^{2+}$-ion intercalation can lead to the superconducting phase transition in bilayer graphene grown on SiC (as previously described).[156] Furthermore, it can also induce the CDW phase transition. For example, Shimizu et al. have intercalated Li and Ca atoms into bilayer graphene grown on SiC, thereby creating Li-intercalated bilayer graphene ($C_6LiC_6$) and Ca-intercalated bilayer graphene ($C_6CaC_6$), respectively.[139] Interestingly, the CDW phase transition can only be observed in $C_6CaC_6$ probably due to its higher carrier density. This is attributed to the fact that Ca donates two electrons while Li offers just one electron. The STM topographic images of bilayer graphene before and after Ca-intercalation can be found in Figures 8M and 8N, respectively. The new pattern in Figure 8N with a long-range modulation is due to the CDW phase transition.[139] Moreover, the CDW phase transition has been corroborated through the analysis of the tunneling spectra under varying conditions (Figure 8O). The gap structure is observed to vanish at 78 K and remains uninfluenced by the magnetic field at 5 K. This indicates that the gap structure can be ascribed to the CDW state, rather than the superconducting state. Moreover, the CDW phase transition is found to be a consequence of the modulation of the long-range potential originating from the SiC substrate.[139] This study highlights the pivotal role of the substrate in triggering CDW phase transitions in graphene and prompts the investigation of whether the CDW phase can be modulated by varying Ca-intercalation levels, i.e., different carrier densities.

Furthermore, intercalation is known to result in the occurrence of CDW phase transitions in TMDs. For example, Hall et al. have intercalated Li atoms into the monolayer 1H-$TaS_2$ on the graphene/Ir(111) substrate.[164] They observed a CDW phase transition in monolayer 1H-$TaS_2$ from a (3 × 3) CDW phase to a (2 × 2) CDW phase. In another example, Chazarin et al. have intercalated Na atoms into bulk $VSe_2$ and observed a CDW phase transition of the surface layer from the initial bulk phase with a periodicity of (4 × 4) to the monolayer phase with a periodicity of ($\sqrt{3} \times \sqrt{7}$).[168] In addition, CDW modulations have been achieved in $MoS_2$ crystals for the first time after the intercalation of $K^+$ ions.[70] With the expansion of the $MoS_2$ lattice, whose vdWs gap has been increased from 6.1 Å to 8.2 Å, two CDW modulations of ($2\sqrt{3} \times 2\sqrt{3}$) R30° and 2 × 2 can be observed in the STM image of $K_{0.4}MoS_2$ (Figure 8P). The former can be attributed to the Fermi surface nesting, while the latter can be ascribed to electron−phonon coupling.[70] This study provides a feasible approach (high-level doping by intercalation) for the generation of CDW phases in other TMDs lacking intrinsic CDW orders.

## 3. Neuromorphic applications

The human brain is composed of $10^{11}$ neurons that are connected to each other via $10^{15}$ synapses. It is capable of rapid processing and storage of information in a parallel way with only a power consumption of approximately 20 watts.[169] Neuromorphic computing represents a novel computing architecture that has been developed with the objective of emulating the rapid and low-power operation of the biological brain. This approach has the great potential to smash energy and speed bottlenecks in traditional von Neumann computing architecture. The construction of large-scale neuromorphic computing systems is made possible through the formation of artificial neural networks comprising a multitude of individual synaptic devices, which are designed to emulate the functions of biological synapses. Neuromorphic devices can be categorized into two groups, namely two-terminal memristors and three-terminal synaptic transistors, according to their architectural characteristics.[170]

Phase transition-enabled resistance switching allows a three-terminal memory transistor to imitate the functions of a biological synapse through the use of multiple controlling terminals, thereby enabling the simulation of synaptic behaviors.[171] Ion intercalation-based electrochemical transistors have been the primary means of demonstrating three-terminal synaptic transistors to date. In this configuration, electrical stimuli can promote intercalation-induced I-M phase transition with nonvolatility across the channels, thereby enabling continuous alteration of the channel conductivity during the phase transition.[74] This dynamic manipulation of channel conductance facilitates the realization of multilevel resistance states. In addition, intercalation-induced ferroelectric and magnetic phase transitions can also result in resistance switching across the channels,[29,54] suggesting their potential applications in multilevel-state-based synaptic devices.

The intercalation-based electrochemical transistor with multiple terminals (at least three terminals) represents an excellent device platform for mimicking biological synapses and implementing complex neuromorphic computing arrays. Firstly, thanks to the small voltage-enabled ion diffusion, the energy consumption of both “write” and “read” operations in the electrochemical transistor is extremely low.[172] Secondly, the dynamic processes of ion absorption (volatile) and intercalation (non-volatile) in the channel can be employed to simulate the short-term plasticity and long-term plasticity of bio-synapses,[171] respectively, enabling computational and memory functions for neuromorphic applications. Thirdly, in contrast to two-terminal memristors, three-terminal synaptic transistors enable concurrent signal processing and learning through the involvement of both the channel and gate,[173] indicating the potential for realizing complex neuromorphic functions.

In general, the channel materials of intercalation-based electrochemical transistors for neuromorphic computing should be semiconducting and readily intercalated. These materials include oxide semiconductors,[77,78,174-177] 2D semiconductors,[178,179] and organic semiconductors.[180,181] The lightest cations, such as protons and $Li^+$ ions, have been inserted to these three types of channels.[77,78,176-181] While $O^{2-}$ anions can also be employed to intercalate into channels of oxide semiconductors, whose resistances are sensitive to the oxide content.[174,175] Other larger ions are extremely difficult to be intercalated into these channels and thus are rarely used in electrochemical transistors for neuromorphic applications.

In the nascent stages of this field, the majority of studies have concentrated on elucidating the operational mechanism and optimizing the performance of single transistors with liquid electrolytes.[75,76,182] However, the fabrication process is incompatible with the standard CMOS technology, as liquid electrolytes are sensitive to solvents and high temperatures. To achieve chip-level applications with synaptic transistors, the CMOS-compatible solid electrolytes have been adopted in recent years.[77,172,183] Currently, researchers are attempting to achieve the sophisticated integration of synaptic transistors, despite the fact that they are still at a low-density level.[77,184,185] Furthermore, novel three-dimensional architectures (e.g., vertical synaptic transistors) are being developed,[172,186,187] which have the potential to facilitate the increase in array density.

### 3.1 Rigid neuromorphic devices

The intercalation-based neuromorphic devices have been demonstrated with both rigid and flexible substrates, which endows them with different application scenarios. For example, intercalation-based flexible devices can be uniquely utilized in wearable electronic systems. Rigid devices, on the other hand, can only be applied in normal in-memory computing systems without deformation, with the aim of achieving high-intensity integration. In general, flexible neuromorphic devices are fabricated on polymer substrates, which are usually incompatible with CMOS processing due to their solubility in commonly used organic solvents and glass transitions at relatively high temperatures. In contrast, rigid neuromorphic devices are always fabricated on conventional silicon substrates, and their fabrication processes could be readily integrated with the CMOS fabrication line. Therefore, rigid neuromorphic devices are a promising avenue for integration into high-density device arrays and could eventually be utilized in wafer-scale integrated memory circuits. See Table 3 for the comparison of typical rigid devices.[77,78,92,174-179,188]

#### 3.1.1 Imitation of biological functions

A variety of intercalation-based neuromorphic devices have been developed with the objective of emulating biological functions such as the plasticity of synapses, the sensory memory of neurons, and classical biological behaviors.[189]

Bao et al., have fabricated dual-gate multiterminal devices based on $MoS_2$ channels and an electrolyte of polyethylene oxide/$LiClO_4$ to emulate a neuron, i.e., to function as a $MoS_2$ neuristor.[190] The top gate is employed to regulate the migration of $Li^+$ ions while the back gate is utilized to modulate the conduction of the channel. The diagrams in Figures 9A and 9B depict the processes of propagating an action potential in a biological nerve axon and the corresponding electrochemical processes mimicked in a $MoS_2$ electrochemical transistor.[190] The intercalation of $Li^+$ ions into the $MoS_2$ channel via the top-gate pulses induces a phase transition, which subsequently reduces the threshold voltage of the back gate. Consequently, an increase in pulse amplitude results in more output spikes. Moreover, the sigmoid relationship between the input voltages and the output spike frequencies in Figure 9C indicates that the $MoS_2$ neuristor can be employed to emulate the artificial neuron in artificial neural networks. In addition, when the top gate is grounded, the device can function as an n-type metal-oxide-semiconductor field-effect transistor (MOSFET); while when the back gate is grounded, the device can mimic a synapse.[190] Therefore, this study proposes a novel methodology for the realization of a multifunctional device, which may facilitate the simplification of circuits for neuromorphic applications.

Similarly, Liu et al. also fabricated a three-terminal synaptic transistor based on $Li^+$-ion absorption and intercalation using a multilayer $MoS_2$ channel and the electrolyte of polyethylene oxide/$LiClO_4$.[178] The transistor has been demonstrated to mimic short-term synaptic plasticity in response to a voltage pulse and long-term synaptic plasticity in the presence of five continuous voltage pulses. Moreover, the intercalation of $Li^+$ ions will simultaneously induce doping and phase transition in the channel.[178] To demonstrate the potential of the device in an attention application, Liu et al. connected it to a feedback loop (Figure 9D), which consisted of a 50 KΩ monitor resistor and a programmed leaky-integrate-and-fire neuron.[178] The membrane potential of the leaky-integrate-and-fire neuron increases when a current flows through it. Once the accumulated potential reaches the threshold voltage of the leaky-integrate-and-fire neuron, it will fire a spike signal, thereby initiating the feedback loop to apply five successive voltage pulses to the gate for the formation of long-term memory.[178] Two typical cases of binary attention have been shown in Figures 9(E) and 9(F), respectively. It can be observed that a high drain-source voltage (0.2 V, input signal "1") and applying gate voltage (2 V, the initial attention signal) correspond to the densest feedback pulses, while a low drain-source voltage (0.1 V, input signal “0”) and without applying gate voltage (0 V, the initial attention signal) correspond to the sparsest feedback pulses.[178]

Accordingly, this novel computing circuit operates as an attention function, wherein the drain-source voltage serves as the input signal, the gate signal serves as the attention weight, and the nonlinear mapping of the input signal serves as the output. The attention mechanism is of significant importance in the deep neural network, as it enables the classification of images and segmentation of scenes.[178] Furthermore, it shows promising applications in the field of automatic driving. It is worth noting that in practical applications, the programmed neuron needs to be replaced by an artificial neuron, e.g., using the $MoS_2$ neuristor above.

Furthermore, intercalation-dependent non-transistor neuromorphic devices have been utilized to mimic biological functions.[80,94,191,192] In these devices, ion migration is driven by lateral electric fields, and the channel resistance is regulated by an intercalation-induced reversible phase transition. The emulation of intricate synaptic behaviors is crucial for artificial neural networks. However, they had not been simulated directly due to the unavailability of suitable devices. To address this issue, Zhu et al. fabricated multi-terminal synaptic devices based on the reversible phase transition of the $MoS_2$ channel between 2H and 1T' induced by Li-ion intercalation, which can be utilized to mimic the competition and cooperation behaviors of synapses.[80]

### 3.1.2 Complex computing

The intercalation-based phase transition has two main applications in neuromorphic computing: pattern/image recognition and logic operations, which can be realized by both rigid memristors and synaptic transistors. Complex computing, such as logic operations, could be more efficient in high-density rigid-device arrays than in flexible devices that are difficult to integrate.

Cui et al. have fabricated a CMOS-compatible electrochemical random-access memory (ECRAM) based on a $H_xWO_3$ channel, a hydrogenated $ZrO_2$ electrolyte, and a H-rich $H_xWO_3$ gate (Figure 9G).[77] The H-rich $H_xWO_3$ layer acts as a proton reservoir, facilitating the supply and extraction of protons. Additionally, it mitigates the built-in potential induced by the different electrochemical potentials of the channel and the gate. The device exhibits favorable retention and perfect symmetry in the potentiation and depression behaviors, and is capable of operating at an ultrahigh speed with a settling time of approximately 13 μs, which is approximately two orders of magnitude faster than that of a typical ECRAM with other intercalants.[77] In addition, the device exhibits remarkable endurance, demonstrating no degradation in performance even after $10^8$-cycle write-read operations. By combining one ECRAM with one silicon MOSFET, it is possible to resolve the issue of write-disturbance. This is because the conduction change of the ECRAM can be ignored once the transistor is switched off, as shown in Figure 9H. As a demonstration, Cui et al. proceeded to fabricate an array of the one transistor-one ECRAM cells, which can execute vector-matrix multiplication parallelly.[77] The array was employed to transform the color of a 4 × 6 pixel image, as illustrated in Figure 9I. The experimental transformation results are close to the software-generated values, which indicates the robust image processing function of the one transistor-one ECRAM cell array. The ECRAM cell array is constructed in a pseudo-crossbar configuration, with the device area being reduced to 150 × 150 $nm^2$, thereby demonstrating the potential for high-density integration.[77] To further enhance the array density, the device architecture might be optimized by referring to the structure of the fin field-effect transistor.

An all-solid vertical neuro-transistor has been successfully fabricated based on an α-$Nb_2O_5$ channel and a $Li_xSiO_2$ electrolyte, as shown in Figure 9J.[172] This special architecture has enabled the realization of short-term memory in an α-$Nb_2O_5$ channel as short as 30 nm, with the read and write energy down to 30 aJ and 275 fJ, respectively. Xu et al. have demonstrated dendritic computing functions in a dual-gate neuro-transistor experimentally.[172] 1. The device is capable of implementing an AND logic function by setting up a threshold current $I_{th}$ (Figure 9K). It is only when the two gates operate simultaneously that the total current exceeds the threshold current, resulting in the device outputting a high signal. Conversely, when the two gates operate separately, the current is always lower than the threshold current, resulting in the device outputting a low signal. 2. The peak channel current and the coincidence degree $A_2/A_1$ decrease with increasing the time interval Δt between the voltage pulses applied to the two gates, as shown in Figure 9L. Therefore, this device is capable of emulating the coincidence detection of two events by biological neurons. Thanks to the compact structure and the high energy efficiency of the vertical neurotransistor, whose fabrication process is compatible with CMOS fabrication processes, it is thus a promising device unit for the realization of integrated neuromorphic computing networks in the future.

### 3.2 Flexible neuromorphic devices

The flexible computing block is a core component of wearable electronic systems. Although the first flexible microprocessor, known as the "PlasticARM", has been developed based on the traditional von Neumann computing architecture, its performance remains significantly inferior to that of the traditional silicon microprocessor because of the limitations of the large thin-film transistors fabricated in a 0.8-μm process.[193] Moreover, the accumulation of heat in general flexible substrates with poor thermal conductivity is more severe than that in silicon substrates. In contrast to the traditional computing architecture, the neuromorphic architecture is capable of executing parallel computing for data processing and storage, which simplifies the required circuits, consumes much less energy, and produces much less heat. Therefore, combining flexible electronics and neuromorphic devices (e.g., the intercalation-based synaptic transistor) to create flexible computing systems represents a significant research direction. It is worth mentioning that flexible neuromorphic devices are not only essential for the advancement of next-generation wearable and implantable computing systems but also have potential applications in soft robots and neuroprosthetics.[194] The performance benchmark of intercalation-based flexible neuromorphic devices can be found in Table 4.[93,171,191,192,195,196]

### 3.2.1 Emulation of biological functions

The tactile and auditory sensing relies on stretchable, soft skin and eardrum, respectively. To build bio-inspired devices and mimic these sensing and storing functionalities, it is essential to adopt flexible substrates. In these devices, strain can be an additional knob to modulate the device conductance, thereby facilitating the realization of multiple functionalities. Recently, a number of biological functions have been emulated by means of intercalation-based flexible devices.

Wei et al. have employed the electrochemical intercalation of $Li^+$ ions in a synaptic transistor on a polyethylene naphthalate (PEN) flexible substrate (Figure 10A).[93] The transistor channel is made of $SnO_2$ nanoparticles (SONPs)-based thin film and a polymethyl methacrylate (PMMA)/electrolyte bilayer. The electrolyte layer is composed of polyethylene oxide/$LiClO_4$, which provides an adequate supply of $Li^+$ ions. The inserted PMMA buffer layer reduces electron traps and the surface energy, thereby optimizing the transistor performance.[93] The employment of two distinct contact modes (i.e., the point and the plane contact) facilitates the diffusion of $Li^+$ ions into the channel, thereby enabling the generation of diverse postsynaptic current behaviors, including varying current gains, paired-pulse facilitation, and memory enhancement. Moreover, the combination of two plane contacts on a single transistor enables the emulation of multiple functions through the use of two-gate inputs, such as parallel information processing and logical operations.[93] Interestingly, this synaptic device is capable of emulating biological functions through the use of two contact modes together. For example, the point input can be used to imitate the food stimulus, while the plane input can be used to imitate the ringing stimulus, thereby enabling the realization of classical Pavlov's learning.[93] Furthermore, Wei et al. verified the stability of the device's performance when subjected to bending. They fabricated device arrays and evaluated the synaptic characteristics when the substrate was bent at a bending radius of 0.8 cm. After bending the substrate for 1000-5000 cycles, they observed that 80% of the device performance was maintained (Figure 10B).[93] In addition, the synaptic device exhibits a response to UV light irradiation. The generation of different quantities of carriers by varying light intensities and durations allows for the manipulation of postsynaptic currents, which can be utilized to emulate pain-perceptual nociceptors.[93] In this study, the point input is achieved through the use of a tungsten probe and it generates intriguing synaptic behaviors. However, it remains a significant challenge to translate this concept into a practical device using microfabrication techniques. Therefore, further studies may be required to obtain the point input by greatly reducing the contact area.

To mimic the biological functions of pain perception and nerve injuries, Lu et al. fabricated a flexible two-terminal memristor (Figure 10C).[192] This device is composed of a conduction channel of 90% semiconducting single-wall carbon nanotube networks, an electrolyte layer of $LiClO_4$ doped polyoxyethylene oxide, and a flexible polyimide substrate. Protons and $Li^+$ ions coexist in the electrolyte and their cooperative behaviors in the synaptic device were employed to emulate the functions of $Na^+$ and $K^+$ in biological systems.[192] The anticlockwise hysteresis in the I-V curves can be attributed to the contributions of protons, while the clockwise hysteresis originates from the contributions of $Li^+$ ions. Based on the different working mechanisms of the device (i.e., proton and $Li^+$ ion hopping at low voltages, and $Li^+$ ion intercalation at high voltages), Lu et al. employed artificial synapses to simulate mild and severe nerve injuries.[192] No injury to the nerve can be imitated at the stimuli of successive low-voltage pulses (i.e., 1 V) with any width, and mild nerve injuries (e.g., neurapraxia in biological systems) can be imitated by applying a high-voltage pulse train (i.e., 3 V) with a width of 10 ms, and the synaptic behavior can be recovered in less than 3 mins (Figure 10D). In contrast, severe nerve injuries (e.g., axonotmesis and neurotmesis in biological synapses) can be imitated at the stimuli of a high-voltage pulse train (i.e., 3 V) with a width of 30 ms, yet the synaptic behavior cannot be recovered in a long time (Figure 10E), resembling the denervation in biological systems. In addition, Lu et al. illustrated the flexibility of the device: almost full device performance is maintained when the device is bent at a small bending radius of approximately 6 mm.[192] Given that proton density is sensitive to environmental humidity and directly affects the electronic properties in this study, a packaged device is preferable in practical applications to maintain humidity and thus proton density. Moreover, this study may provide a potential solution to the human-machine interface.

### 3.2.2 Image recognition

Flexible-device-based neuromorphic computing is a crucial component of next-generation wearable computing systems, which extend beyond health monitoring to encompass applications such as robots and brain-computer interfaces.[194] Moreover, flexible neuromorphic devices can be integrated with flexible nanogenerators to create self-powered computing systems.[197] In addition, intercalation-based electrochemical transistors are capable of concurrent signal processing and learning.[173] This necessitates the exploration of intercalation-based flexible neuromorphic devices, with which image recognition has been the primary computing application recently.

Hwang et al. developed a novel method for fabricating synaptic devices on flexible substrates (colorless polyimide) (Figure 10F):[195] a $MoS_2$ layer is directly synthesized on a polymer substrate via plasma-enhanced chemical vapor deposition at a relatively low temperature (200 °C) to prevent substrate degradation. Exceptionally, a $LiSiO_x$ electrolyte layer is deposited directly on the $MoS_2$ channel via co-sputtering.[195] The advantage of this method is that it circumvents the conventional transfer process for $MoS_2$; consequently, the transfer-induced defects and associated time consumption can be circumvented. Therefore, this method is applicable to the fabrication of other similar flexible neuromorphic devices and is promising for practical use in semiconductor processing lines. The synaptic device, which imitates the transmission of bio-signals between two neurons, is based on the process of intercalating/de-intercalating Li ions at positive/negative gate voltages.[195] The electrochemical intercalation process is reflected by the anticlockwise hysteresis of the transfer curve due to the nonvolatile properties. As displayed in Figure 10G, the nonlinearity and the asymmetry ratio are extracted to be 0.55 and 0.22, respectively, indicating excellent neuromorphic synaptic characteristics. Hwang et al. evaluated the flexibility of the synaptic device by attaching it to a semi-cylindrical support with varying diameters.[195] The device exhibited stable and repeatable endurance characteristics under four different strains after 400 cycles, indicating that it can maintain good synaptic performance when subjected to strain up to 0.25%. Moreover, the durability of the synaptic device was further verified by repeatedly bending the device for 700 cycles using a bending machine. Additionally, Hwang et al. conducted simulations of handwritten digit recognition using a multilayer neural network structure with parameters extracted from the synaptic transistor in various bend states (Figure 10H).[195] The accuracy of the device is 95.2% in the flat state, but decreases to 94.3% and 92.4% at bend diameters of 30 mm and 20 mm, respectively. Moreover, the accuracy remains at a high value of 90% even after 300 bending cycles. The remarkable flexibility of this synaptic transistor renders it an attractive option for incorporation in flexible neuromorphic computing systems.

Deng et al. fabricated a Mott synaptic transistor on a flexible mica substrate based on a $VO_2$ channel and a solid electrolyte (Figure 10I).[171] The $VO_2$ film was deposited on mica by pulsed laser deposition, and the electrolyte gel was prepared by mixing an ionic liquid and a copolymer.[171] When the gate voltage is low, insufficient to activate the electrochemical reaction, the organic ions of the electrolyte only migrate and then accumulate at the interface between the $VO_2$ channel and the electrolyte, forming an electric double layer. After the withdrawal of $V_G$, the ions at the interface return to the electrolyte, resulting in a rapid recovery of the channel resistance and exhibiting volatile behavior.[171] Upon reaching the electrochemical reaction threshold, the gate voltage induces the intercalation of protons into the $VO_2$ channel, leading to the formation of a stable metallic $H_xVO_2$ phase, which exhibits nonvolatile behaviors.[171] Accordingly, the synaptic transistor's behavior at low and high gate voltages is employed to imitate the short-term and long-term synaptic plasticity, respectively. Moreover, the synaptic characteristics of the device demonstrate remarkable stability and reliability when subjected to bending deformation. The potentiation and depression properties remain largely unaltered even after 500-cycle bending at the bending radius of 6 cm, as shown in Figure 10J.[171] In addition, a convolutional neural network has been developed to assess the computing capability of the synaptic transistor, and the recognition accuracy reaches 95%, approaching the ideal value (Figure 10K).[171] Therefore, this flexible Mott synaptic transistor shows considerable potential for deployment in wearable neuromorphic applications, due to its relatively low energy consumption ($8.8 \times 10^{-13}$ J for each synaptic event) and high stability and reliability under bending.

## CONCLUSIONS AND PERSPECTIVES

The intercalation technique has been demonstrated to be a powerful technique to expand the material family, manipulate material properties, and stimulate the development of novel devices. In this review, we have summarized the fundamental physics of intercalation-based phase transitions and their applications in emerging brain-inspired computing paradigms. Additionally, we outline the involved intercalation and characterization techniques. Despite the considerable advancements made in functional material intercalation, several pivotal challenges remain, accompanied by exciting opportunities (Figure 11).

Firstly, from a material perspective, several issues need to be addressed. For example, it is challenging to achieve homogeneous intercalation and uniform phase transition, thus leading to disparate properties across different areas and poor device performance. Moreover, compounds intercalated with alkali metals and organic molecules are sensitive to air and elevated temperatures, respectively, because of the inherent instability of the intercalants. Therefore, they typically degrade under ambient conditions or during device fabrication. Furthermore, in-situ techniques with high accuracy are required to monitor intercalation-based phase transitions in nano-devices. Finally, the majority of intercalation methods developed are targeted for laboratory use only, necessitating the pressing need for the development of a simple, cost-effective, and scalable intercalation method for industrial applications in the future.

Secondly, from a physics perspective, the intercalation mechanism remains unclear due to the involvement of numerous factors, such as strain and charge doping. The intercalation process is a collective interaction among ion migration, charge transfer, and lattice distortion, which presents a significant challenge for theoretical and experimental investigation. Furthermore, theoretical studies that provide more guidance for the design of intercalated compounds with desired properties can significantly accelerate experimental development toward uncharted physics. The intercalation of low-dimensional functional or strongly correlated materials to achieve the coexistence of mutually exclusive phenomena, such as ferroelectricity and superconductivity, represents a highly appealing avenue of research. In addition, the migration trajectory of inserted ions (e.g., protons) and their interplay with long-range dipole interaction in ferroelectric channel materials remains largely uninvestigated.

Finally, from a device perspective, the intercalation/deintercalation rate represents a crucial factor determining the operational speed in most neuromorphic applications. To achieve ultrafast switching in an electrochemical synaptic device, the intercalation/deintercalation process must be accelerated. Towards this goal, it is necessary to enhance the ionic conductivity of the electrolyte, the interfacial exchange current density, and the ion diffusivity. Theoretically, an optimal electrochemical transistor for neuromorphic computing exhibits a low operation voltage, a short programming pulse, and a large on/off ratio. The low operation voltage and the short programming pulse make the device energy-efficient, while the large on/off ratio can correspond to a multitude of analog states. Nevertheless, it remains a significant challenge to achieve these characteristics in a single intercalation-based device. Consequently, it is essential to identify suitable channel and electrolyte materials, while simultaneously optimizing the device architecture. Considering the array integration in practical applications, a solid electrolyte that is compatible with the CMOS process is favorable. Furthermore, the cycle-to-cycle and device-to-device variations must be reduced in order to achieve the high performance of integrated arrays. In addition, intercalation can be incorporated into innovative devices comprising diverse heterostructures, layered multiferroic materials, and sliding ferroelectric materials, which may motivate the exploration of exotic properties and device applications.

## ACKNOWLEDGMENTS

This research was supported by the National Natural Science Foundation of China (No. 62304202) and the Zhejiang Provincial Natural Science Foundation of China (grant no. LDT23F04013F04).

## AUTHOR CONTRIBUTIONS

X.H. and F.X. co-wrote the paper. F.X. designed and supervised the project. All the authors contributed to the revision of the manuscript.

## DECLARATION OF INTERESTS

The authors declare no competing interests.

**Table 1. Comparison of different intercalation methods**

| Method | Advantage | Disadvantage | Application |
|---|---|---|---|
| Liquid-phase intercalation | • Easy to proceed<br>• Scalable<br>• Reversible<br>• High level of intercalation<br>• Zero-valent metal intercalation<br>• Applicable to all types of hosts<br>• Clean products | • Difficult to precisely control the intercalation<br>• Difficult to perform in-situ characterization<br>• Relatively low intercalation rate | • Search for new materials<br>• Exfoliation of bulk hosts<br>• Modify the physical or chemical characteristics |
| Electrochemical intercalation | • Scalable<br>• High degree of controllability<br>• High degree of reversibility<br>• High intercalation speed<br>• Allow in-situ characterization | • Ionic intercalants only<br>• Conductive hosts only | • Search for new materials and new phenomena<br>• Exfoliation of bulk hosts<br>• Manipulate the physical or chemical characteristics<br>• Energy storage devices<br>• Memory devices |
| Vapor-phase intercalation | • Scalable<br>• Applicable to many intercalant-host pairs<br>• High level of intercalation | • High temperature<br>• Vacuum/inert atmosphere<br>• Nonreversible<br>• Low degree of controllability<br>• Difficult to perform in-situ characterization | • Search for new materials<br>• Modify the physical or chemical characteristics |
| Solid-phase intercalation | • Easy to proceed | • Low intercalation rate<br>• Inhomogeneous intercalation<br>• Nonreversible | • Search for new materials |

**Table 2. Comparison of various characterization techniques for intercalation**

| Technique | Advantage | Disadvantage |
|---|---|---|
| TEM | • Characterization of intercalated structures at the atomic scale<br>• Provide lattice parameters of intercalated crystals via SAED<br>• Mapping of intercalated elements via EDS or EELS at high resolution | • Difficult to prepare samples<br>• Difficult to perform in-situ characterization for intercalation<br>• High-energy electron beams may damage samples<br>• High vacuum is needed |
| STM | • Atomic resolution of the surface structure after intercalation<br>• Provide information about intercalation-induced change in the electronic band gap via STS<br>• Allow in-situ characterization for intercalation | • Only provide the surface information after intercalation<br>• Samples should be conductive<br>• Imaging quality could be degraded by the residual particles on the surface after intercalation |
| AFM | • Easy to prepare samples<br>• Sensitive to intercalation-induced variation in thickness<br>• Samples can be insulating<br>• Allow in-situ characterization for intercalation | • The lateral resolution is decided by the probe radius<br>• Artifacts may be introduced by a probe with a large radius<br>• The probe working in the contact mode may damage samples |
| XRD | • Easy to prepare samples<br>• Nondestructive<br>• Precise analysis of intercalation-induced changes related to structures, sizes, strains<br>• Applicable to various types of intercalated materials<br>• Allow in-situ characterization for intercalation | • Need a large quantity of the intercalated samples<br>• Weak signals for light elements in hosts and intercalants |
| XPS | • Nondestructive<br>• Sensitive to intercalation-induced change in binding energies<br>• Quantitative analysis of alterations in chemical states after intercalation | • Only provide the surface information after intercalation<br>• Incapable of analyzing H and He elements<br>• Low lateral resolution<br>• High vacuum is needed |
| Raman spectroscopy | • Fast<br>• Nondestructive<br>• Easy to prepare samples<br>• Sensitive to intercalation<br>• Relatively high resolution to characterize intercalation<br>• Need a small quantity of the intercalated samples<br>• Allow in-situ characterization for intercalation | • Fluorescent interference<br>• Weak signals for atomically thin samples |

TEM, transmission electron microscope; SAED, selected area electron diffraction; EDS, energy-dispersive X-ray spectroscopy; EELS, electron energy loss spectroscopy; STM, scanning tunneling microscope; STS, scanning tunneling spectroscopy; AFM, atomic force microscope; XRD, X-ray diffraction; XPS, X-ray photoelectron spectroscopy.

| Table 3. Benchmark of rigid neuromorphic devices | | | | | | | | | |
|---|---|---|---|---|---|---|---|---|---|
| Material channel (electrolyte) | Dimension length×width | Minimum pulse width | Writing voltage | On/off ratio | Analog state | Linearity | Endurance | Retention | Reference |
| **$H^+$-ion intercalation** | | | | | | | | | |
| $WO_3$ ($ZrO_2$) | 10 μm × 3 μm | 5 μs | ±4 V | ~21 | 64 | – | $10^8$ | ~4000 s | Cui et al.[77] |
| $Ti_3C_2T_x$ ($H_2SO_4$-PVA) | 1000 μm × 20 μm | 200 ns | ±1 V | ~6 | 50 | 0.65 /1.59 | $10^8$ | 5 min | Melianas et al.[188] |
| p(g2T-TT) (EMIM:TFSI PVDF-HFP) | 45 μm × 15 μm | 20 ns | ±1 V | ~4.7 | 100 | – | $2.1×10^9$ | >5 min | Melianas et al.[181] |
| **$Li^+$-ion intercalation** | | | | | | | | | |
| $Li_xTiO_2$ ($LiClO_4$-PEO) | 10 μm × 8 μm | 10 ms | ±300 mV | ~1.9 | 250 | – | $10^6$ | >7 h | Li et al.[176] |
| α-$Nb_2O_5$ ($Li_xSiO_2$) | 1 μm × 50 μm | 0.5 ms | ±3 V | ~2.7 | 20 | 0.632 /0.976 | 200 | 1000 s | Xu et al.[177] |
| Graphene ($LiClO_4$-PEO) | 12 μm × 3 μm | 10 ms | ±50 pA | ~9.2 | 250 | – | 500 | 13 h | Sharbati et al.[92] |
| $MoS_2$ ($LiClO_4$-PEO) | ~1 μm × – | 200 ms | +0.8 V /−0.4 V | ~2.1 | 100 | – | 800 | – | Liu et al.[178] |
| $MoS_2$ ($LiSiO_x$) | 60 μm × 100 μm | 1 s | +5.5 V /−5 V | ~4.1 | 20 | −0.35 /−2.35 | 560 | 500 s | Park et al.[179] |
| **$O^{2-}$-ion intercalation** | | | | | | | | | |
| $WO_3$ (YSZ) | 5 μm × 10 μm | 10 ms | ±1 V | 1.17 | 100 | 1.6 /0.25 | $10^3$ | ~$6×10^3$ s | Nikam et al.[174] |
| PCMO ($HfO_x$) | 20 μm × 50 μm | 1 s | +3 V /−3.75 V | 40 | 100 | 0.58 /−0.86 | 4000 | 1000 s | Lee et al.[175] |

PVA, poly(vinyl alcohol); p(g2T-TT), poly(2-(3,3-bis(2-(2-(2-methoxyethoxy)ethoxy)ethoxy)-[2,2-bithiophen]-5-yl)thieno[3,2-b]thiophene); EMIM:TFSI, 1-ethyl-3-methylimidazolium bis(trifluoromethylsulfonyl)imide; PVDF-HFP, poly(vinylidene fluoride-co-hexafluoropropylene); PEO, polyethylene oxide; YSZ, $Y_2O_3$-stabilized $ZrO_2$; PCMO, $Pr_{0.7}Ca_{0.3}MnO_3$.

**Table 4. Benchmark of flexible neuromorphic devices**

| Intercalant | Channel | Electrolyte | Substrate | Dimension | Bending radius | Bending cycle | Neuromorphic function | Reference |
|---|---|---|---|---|---|---|---|---|
| $H^+$ and $Li^+$ | s-SWCNTs | $LiClO_4$-PEO | PI | – | 6 mm | – | Emulation of pain perception and nerve injuries | Lu et al.[192] |
| $Li^+$ or $K^+$ | $MoS_2$ | – | PDMS | – | 10 mm | 100 | LTP and LTD | Lee et al.[191] |
| $Li^+$ | $MoS_2$ | $LiSiO_x$ | cPI | ~50 μm × 100 μm | 10 mm | 700 | LTP, LTD, and image recognition | Hwang et al.[195] |
| $[TFSA]^-$ | P3HT | [EMI][TFSA]-P(VDF-HFP) | PET | ~100 μm × 1000 μm | 0.5 cm | – | EPSC, PPF, supralinear integration, spatiotemporal integration, and image recognition | Liu et al.[196] |
| $Li^+$ | SONP | $LiClO_4$-PEO | PEN | 100 μm × 2000 μm × 11 (channel) | 0.8 cm | 10000 | EPSC, PPF, spatiotemporal learning rules, logic operations, classical conditioning behavior, and memory enhancement | Wei et al.[93] |
| $H^+$ | $VO_2$ | [DEME][$BF_4$]-(PVDF-HFP) | Mica | 0.56 $mm^2$ | 6 mm | 500 | EPSC, PPF, LTP, LTD, SRDP, STDP, emulation of nociceptor, and image recognition | Deng et al.[171] |

s-SWCNTs, semiconducting single-wall carbon nanotubes; PEO, polyethylene oxide; PI, polyimide; PDMS, poly(dimethylsiloxane); LTP, long-term potentiation; LTD, long-term depression; cPI, colorless polyimide; TFSA, bis(trifluoromethylsulfonyl)amide; P3HT, poly(3-hexylthiophene); [EMI][TFSA], 1-ethyl-3-methylimidazolium bis(trifluoromethylsulfonyl)amide; P(VDF-HFP), poly(vinylidene fluoride-co-hexafluoropropylene); PET, polyethylene terephthalate; EPSC, excitatory postsynaptic current; PPF, paired pulse facilitation; SONP, $SnO_2$ nanoparticle; PEN, polyethylene naphthalate; [DEME][$BF_4$], N, N-diethyl-N-methyl-N-(2-methoxyethyl) ammonium tetrafluoroborate; PVDF-HFP, poly(vinylidene fluoride-hexafluoropropylene); SRDP, spike rate-dependent plasticity; STDP, spike timing-dependent plasticity.

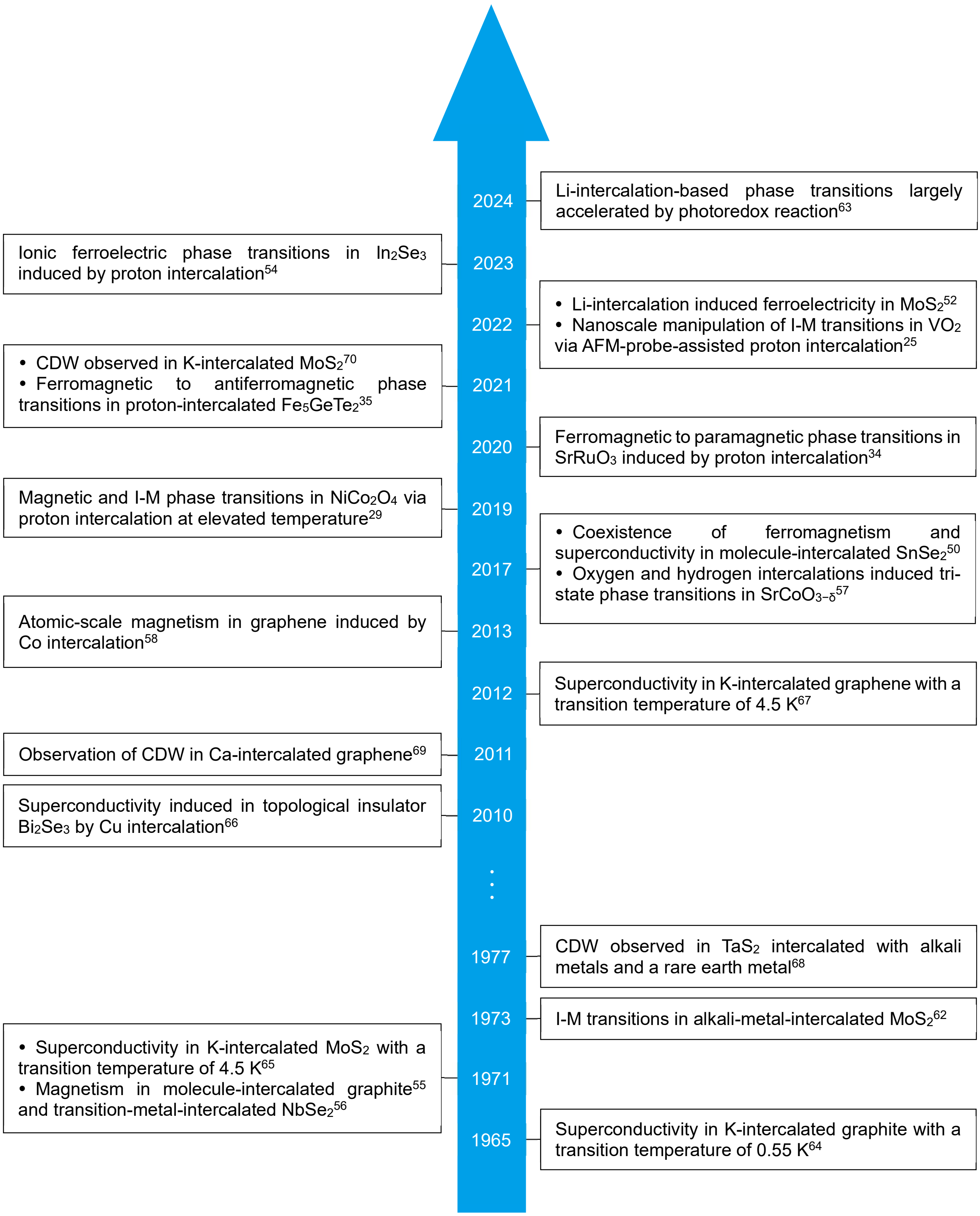

**Figure 1. Timeline of key landmarks for exploring intercalation-based phase transitions.**

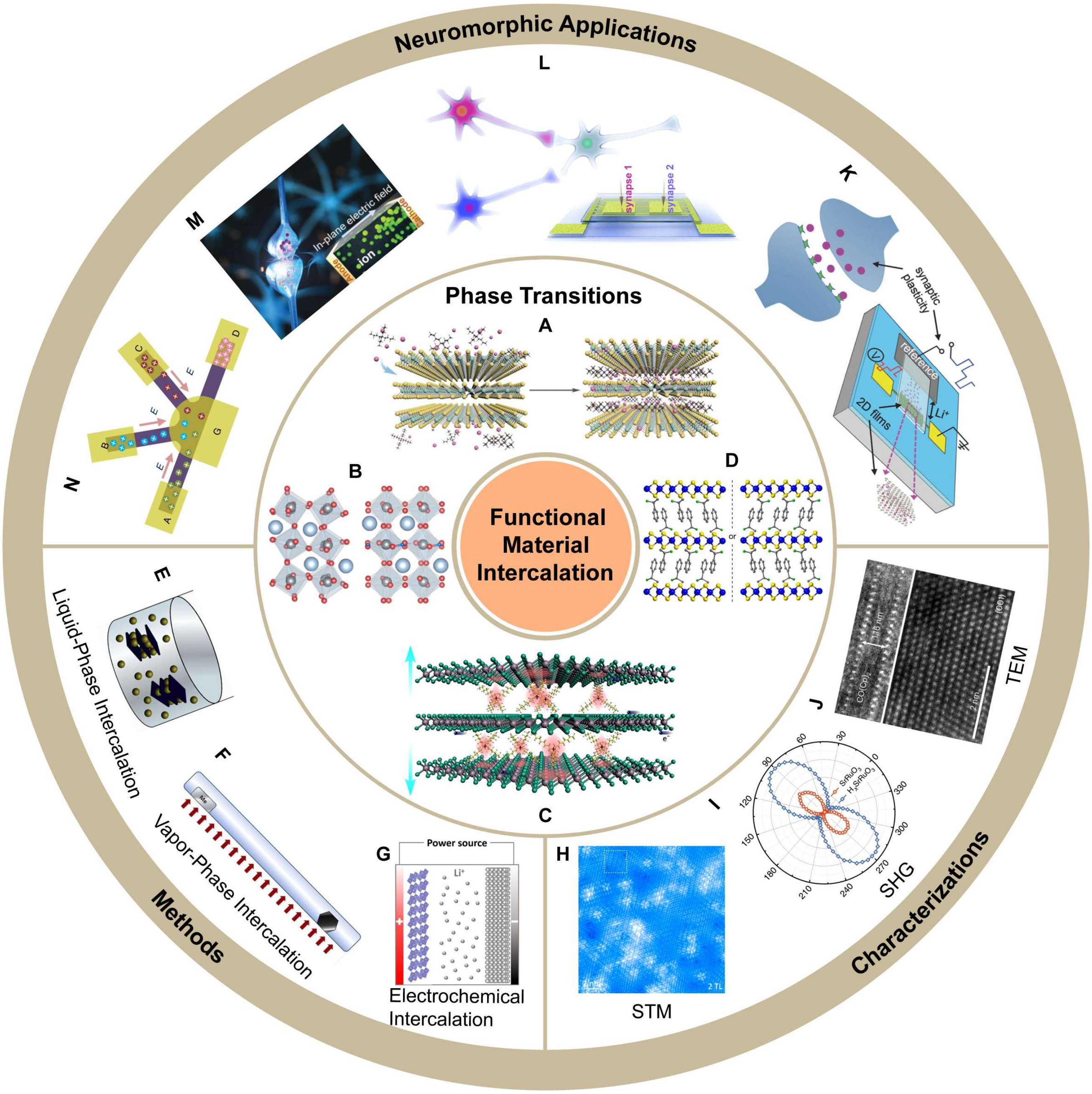

**Figure 2. Schematic diagram of functional material intercalation.**

(A-D) Intercalation-based phase transitions. (A) is adapted with permission from Li et al.[51] Copyright 2020, WILEY-VCH Verlag GmbH & Co. KGaA, Weinheim. (B) is adapted with permission from Shen et al.[86] Copyright 2021, The Author(s). (C) is reproduced with permission from Zhao et al.[87] Copyright 2021, American Chemical Society. (D) is adapted with permission from Qian et al.[88] Copyright 2022, The Author(s), under exclusive license to Springer Nature Limited.

(E-G) Intercalation methods. (E) is adapted with permission from Liu et al.[89] Copyright 2013, Elsevier B.V. (F) is adapted with permission from Nikonov et al.[90] Copyright 2017, Royal Society of Chemistry. (G) is adapted with permission from He et al.[91] Copyright 2023, American Chemical Society.

(H-J) Characterization techniques. (H) is reproduced with permission from Ren et al.[72] Copyright 2022, American Chemical Society. (I) is adapted with permission from Li et al.[34] Copyright 2020, The Author(s). (J) is adapted with permission from Liu et al.[31] Copyright 2024, The Author(s), under exclusive license to Springer Nature Limited.

(K-N) Neuromorphic applications. (K) is adapted with permission from Sharbati et al.[92] Copyright 2018, WILEY-VCH Verlag GmbH & Co. KGaA, Weinheim. (L) is adapted with permission from Wei et al.[93] Copyright 2020, Elsevier Ltd. (M) is adapted with permission from Qin et al.[94] Copyright 2022, Wiley-VCH GmbH. (N) is adapted with permission from Zhu et al.[80] Copyright 2018, The Author(s), under exclusive license to Springer Nature Limited.

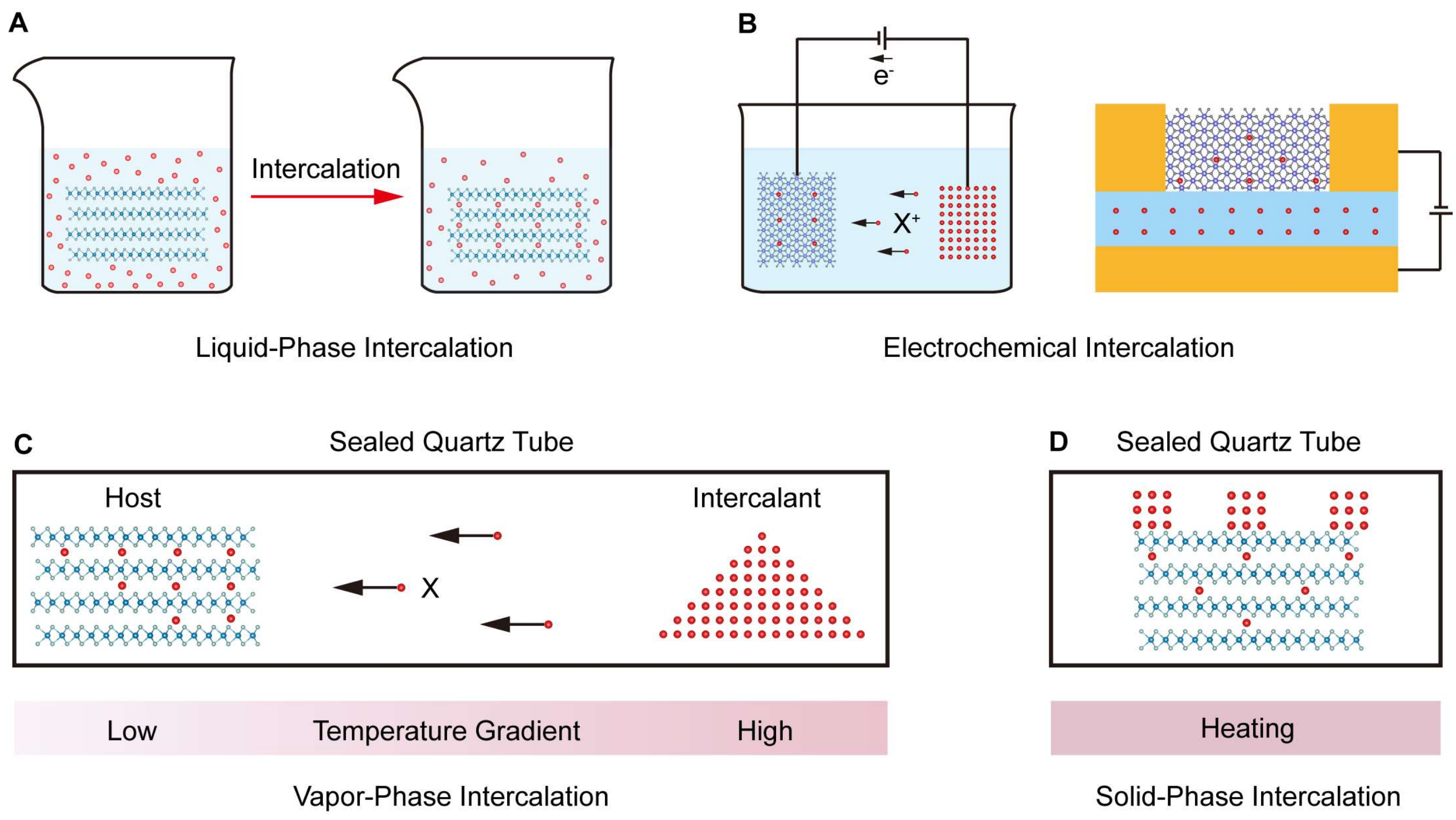

**Figure 3. Schematic illustration of various intercalation methods.**

(A) Liquid-phase intercalation.

(B) Electrochemical intercalation with a liquid/solid electrolyte.

(C) Two-zone vapor-phase intercalation.

(D) Solid-phase intercalation.

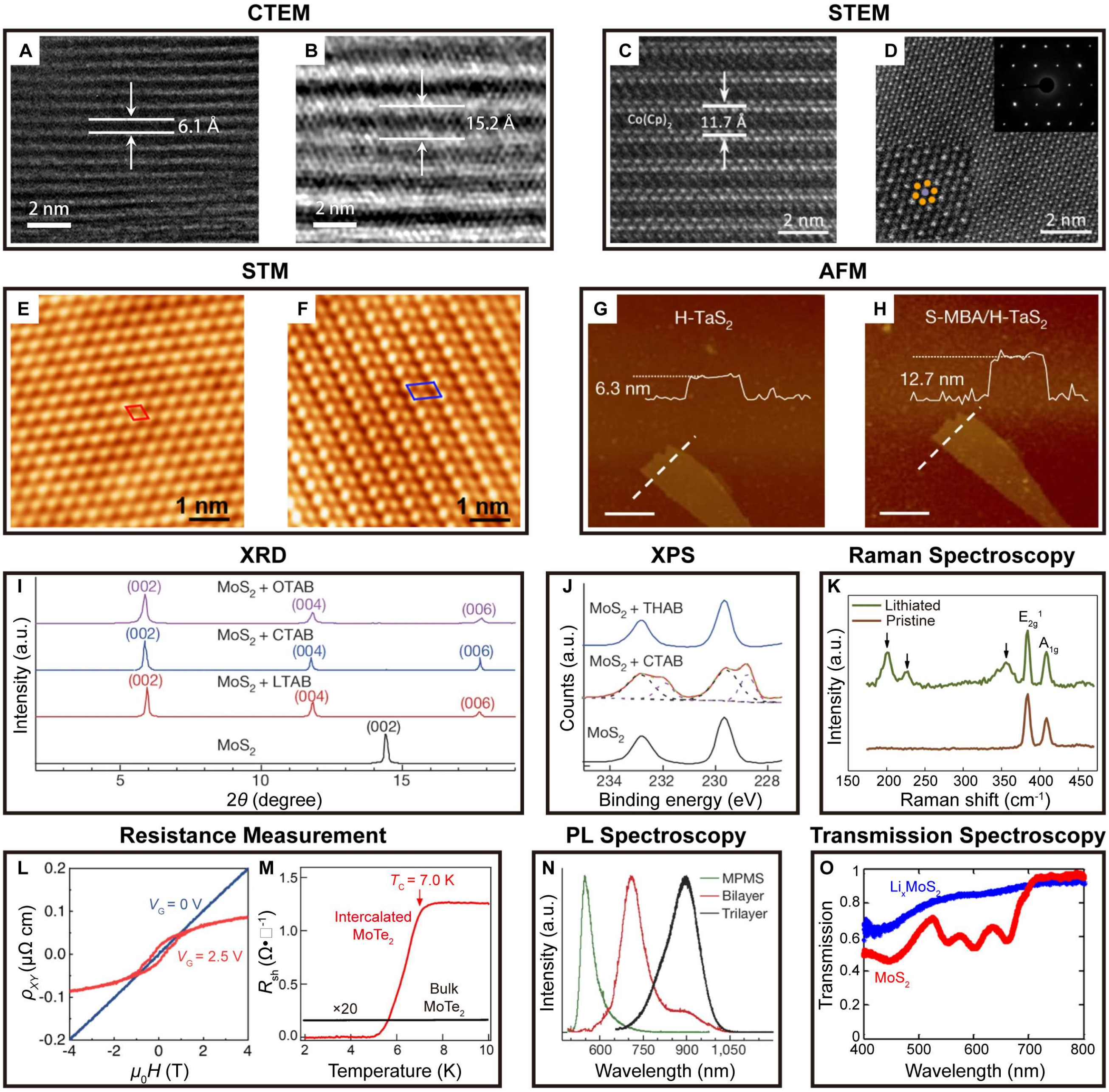

**Figure 4. Common characterization techniques for intercalation.**

(A and B) Cross-sectional conventional transmission electron microscopy (CTEM) images of $SnSe_2$ before (A) and after (B) the intercalation of tetraoctylammonium cations. Adapted with permission from Zhao et al.[87] Copyright 2021, American Chemical Society.

(C and D) Cross-sectional (C) and plane-view (D) scanning transmission electron microscopy (STEM) images of a $SnSe_2$−$Co(Cp)_2$ superlattice. Adapted with permission from Li et al.[50] Copyright 2017, American Chemical Society.

(E and F) Atomic-resolution scanning tunneling microscopy (STM) images of pristine 1T-$NiTe_2$ (E) and self-intercalated $Ni_3Te_4$ (F). Adapted with permission from Pan et al.[137] Copyright 2022, American Chemical Society.

(G and H) Atomic force microscopy (AFM) images of H-$TaS_2$ before (G) and after (H) intercalation of chiral molecules S-α-methylbenzylamine (S-MBA) (scale bars: 500 nm). Adapted with permission from Qian et al.[88] Copyright 2022, The Author(s), under exclusive license to Springer Nature Limited.

(I) X-ray diffraction (XRD) patterns of $MoS_2$ and $MoS_2$ intercalated with lauryltrimethylammonium bromide (LTAB), cetyltrimethylammonium bromide (CTAB), and octadecyltrimethylammonium bromide (OTAB), respectively. Adapted with permission from Wang et al.[125] Copyright 2018, Macmillan Publishers Limited, part of Springer Nature.

(J) X-ray photoelectron spectroscopy (XPS) spectra of pristine $MoS_2$, $MoS_2$ intercalated with CTAB, and $MoS_2$ intercalated with tetraheptylammonium bromide (THAB). Adapted with permission from Wang et al.[125] Copyright 2018, Macmillan Publishers Limited, part of Springer Nature.

(K) Raman spectra of $MoS_2$ and Li-intercalated $MoS_2$. Adapted with permission from Zhu et al.[80] Copyright 2018, The Author(s), under exclusive license to Springer Nature Limited.

(L) Magnetic-field-dependent Hall resistivity of pristine ($V_G$ = 0 V) and protonated ($V_G$ = 2.5 V) $CaRuO_3$. Adapted with permission from Shen et al.[86] Copyright 2021, The Author(s).

(M) Temperature-dependent sheet resistance of organic-cation-intercalated $MoTe_2$ with a superconducting transition temperature of 7.0 K. Adapted with permission from Zhang et al.[121] Copyright 2019, Science China Press.

(N) Photoluminescence (PL) spectra of black phosphorus at different intercalation stages. Adapted with permission from Wang et al.[125] Copyright 2018, Macmillan Publishers Limited, part of Springer Nature.

(O) Optical transmission spectra of pristine and Li-intercalated $MoS_2$. Adapted with permission from Xiong et al.[49] Copyright 2015, American Chemical Society.

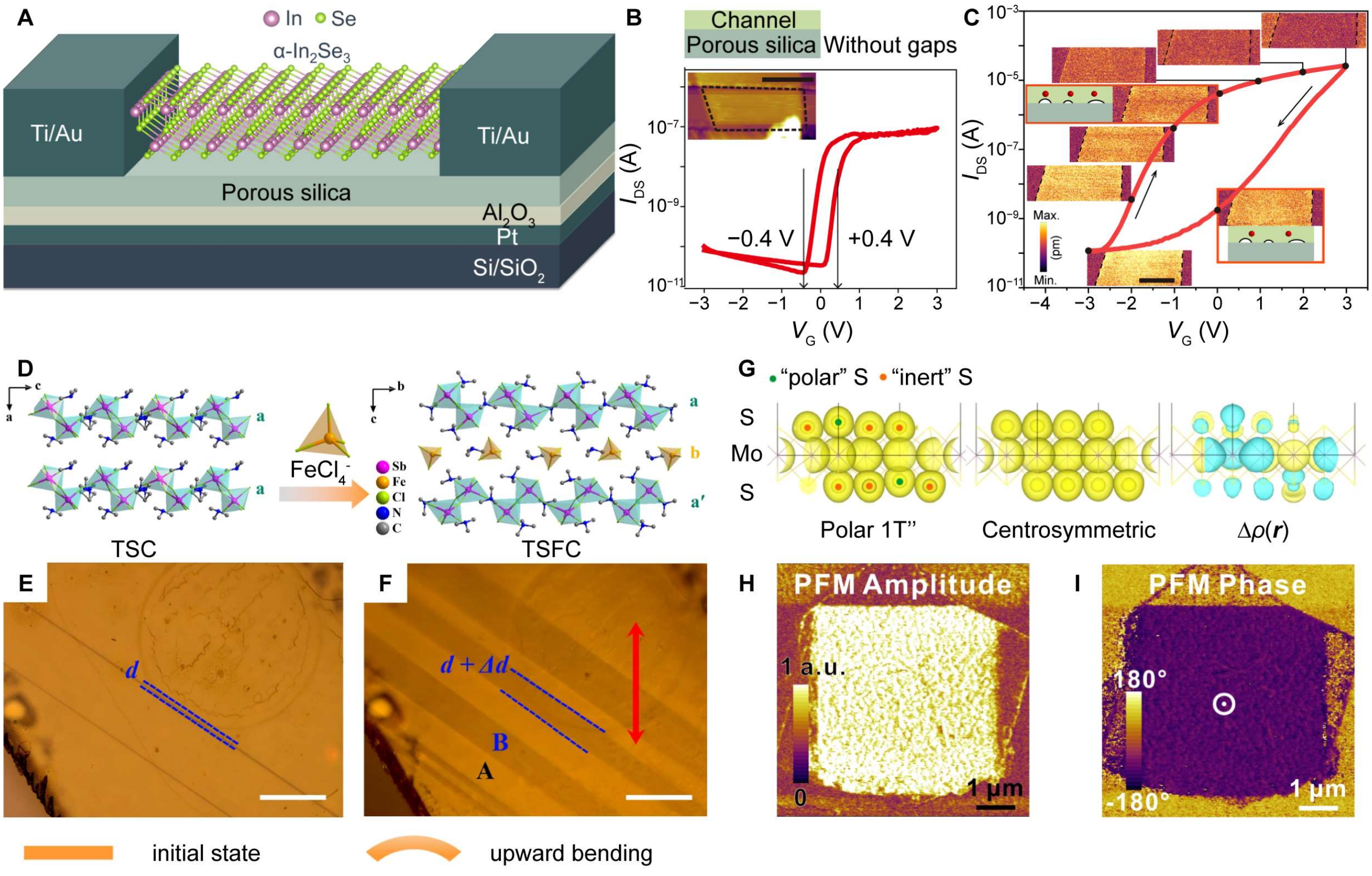

**Figure 5. Ferroelectric phase transitions.**

(A) Schematics of an α-$In_2Se_3$ electrochemical transistor, with a porous silica layer supplying or storing protons. Adapted with permission from He et al.[54] Copyright 2023, The American Association for the Advancement of Science.

(B) Transfer curve of a typical α-$In_2Se_3$ electrochemical transistor (illustrated in (A)) without gaps at the α-$In_2Se_3$/porous silica interface (scale bar: 1 μm). Adapted with permission from He et al.[54] Copyright 2023, The American Association for the Advancement of Science.

(C) Transfer curve of a typical α-$In_2Se_3$ electrochemical transistor (illustrated in (A)) with gaps at the α-$In_2Se_3$/porous silica interface (scale bar: 2 μm). Adapted with permission from He et al.[54] Copyright 2023, The American Association for the Advancement of Science.

(D) Structural transition from $(TMA)_3Sb_2Cl_9$ (TSC) to $(TMA)_4$-Fe(iii)$Cl_4$-$Sb_2Cl_9$ (TSFC) (TMA represents trimethylammonium) after intercalation of $FeCl_4^-$ ions. Adapted with permission from Wu et al.[53] Copyright 2022, The Author(s).

(E and F) Polarization light images highlighting the domains in pristine (E) and upward-bended (F) TSFC (scale bars: 50 μm). Adapted with permission from Wu et al.[53] Copyright 2022, The Author(s).

(G) Calculated electronic charge density for the polar 1T”-$MoS_2$ structure (left panel) and a reference centrosymmetric structure (center panel), and the charge density difference between them (right panel). Adapted with permission from Lipatov et al.[52] Copyright 2022, The Author(s).

(H and I) PFM amplitude (H) and phase (I) images of a typical 1T”-$MoS_2$ flake after mechanical poling. Reproduced with permission from Lipatov et al.[52] Copyright 2022, The Author(s).

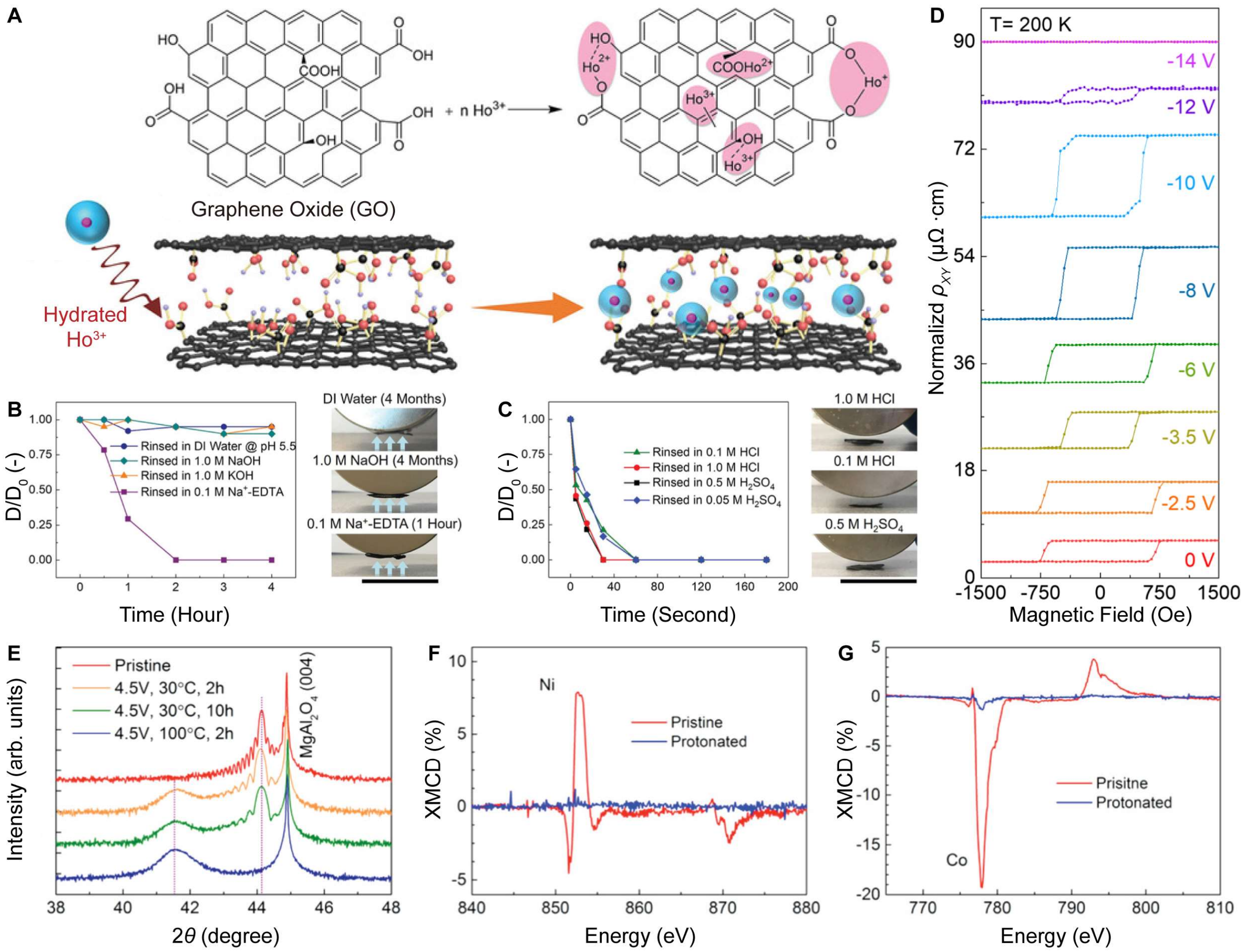

**Figure 6. Magnetic phase transitions.**

(A) Schematic illustration of intercalation of hydrated $Ho^{3+}$ ions into graphene oxide. Adapted with permission from Li et al.[32] Copyright 2019, WILEY-VCH Verlag GmbH & Co. KGaA, Weinheim.

(B and C) Stability tests of the magnetic response of $Ho^{3+}$-GO films rinsed in neutral and basic (B) and acidic (C) solutions for different lengths of time. $D_0$ and D are the magnetically responsive distances before and after the stability tests. Scale bars: 3 cm. Adapted with permission from Li et al.[32] Copyright 2019, WILEY-VCH Verlag GmbH & Co. KGaA, Weinheim.

(D) Evolution of anomalous Hall effect hysteresis loop of $Cr_{1.2}Te_2$ under different protonic gate voltages. Adapted with permission from Tan et al.[33] Copyright 2023, American Physical Society.

(E) $2\theta$–$\omega$ scans of pristine and gated $NiCo_2O_4$ films grown on $MgAl_2O_4$ (004) substrates under different conditions. Adapted with permission from Wang et al.[29] Copyright 2019, WILEY-VCH Verlag GmbH & Co. KGaA, Weinheim.

(F and G) X-ray magnetic circular dichroism (XMCD) spectra of pristine and protonated $NiCo_2O_4$ obtained at the nickel (F) and the cobalt (G) *L*-edges. Adapted with permission from Wang et al.[29] Copyright 2019, WILEY-VCH Verlag GmbH & Co. KGaA, Weinheim.

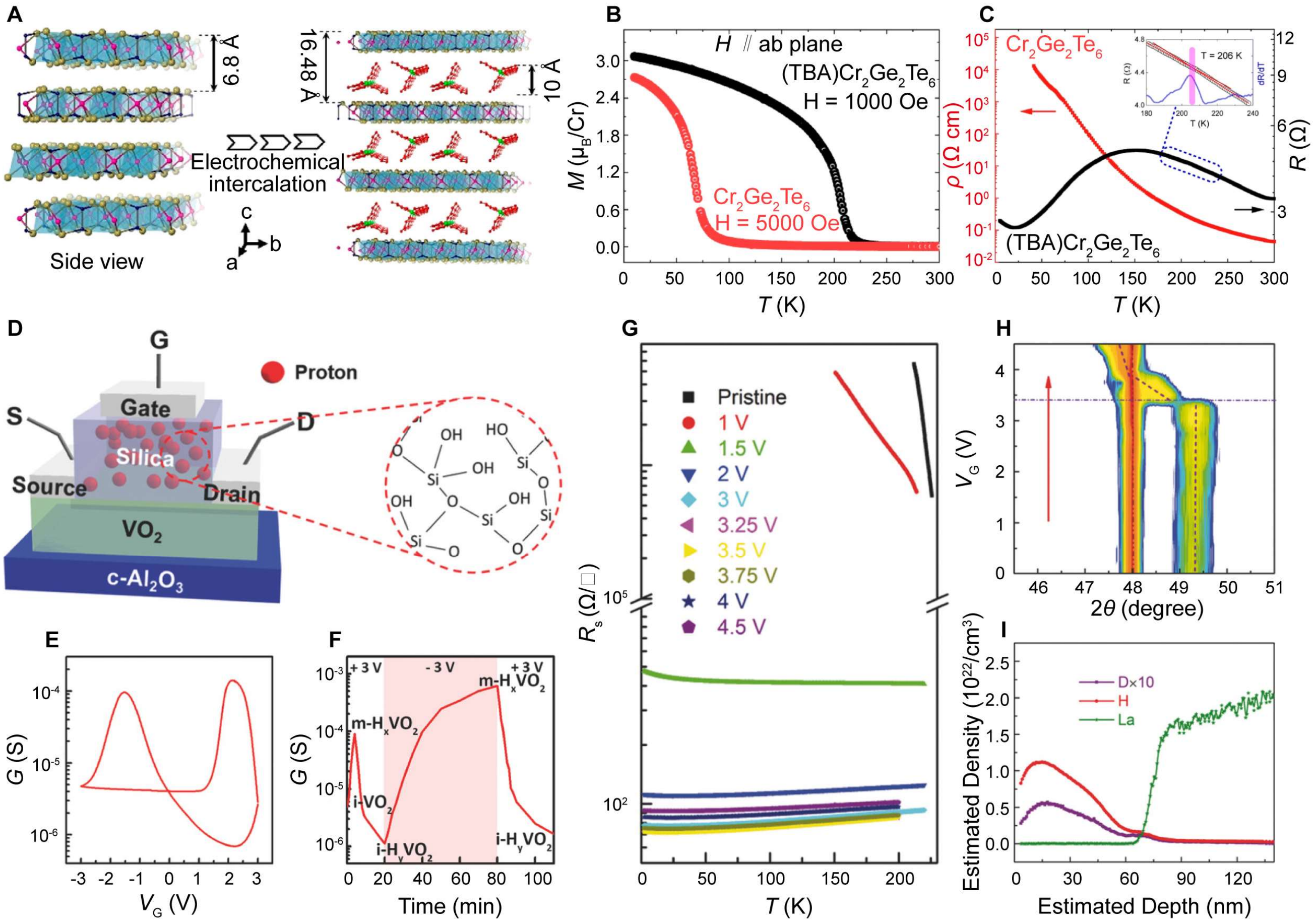

**Figure 7. I-M phase transitions.**

(A) Schematic illustration of intercalation of tetrabutyl ammonium ($TBA^+$) into $Cr_2Ge_2Te_6$. Adapted with permission from Wang et al.[59] Copyright 2019, American Chemical Society.

(B) Temperature-dependent magnetic susceptibility of $Cr_2Ge_2Te_6$ and (TBA)$Cr_2Ge_2Te_6$. Adapted with permission from Wang et al.[59] Copyright 2019, American Chemical Society.

(C) Resistivity-temperature curve for $Cr_2Ge_2Te_6$ and resistance-temperature curve for (TBA)$Cr_2Ge_2Te_6$. Adapted with permission from Wang et al.[59] Copyright 2019, American Chemical Society.

(D) Schematic diagram of a $VO_2$ electrochemical transistor with silica as the proton reservoir. Reproduced with permission from Jo et al.[26] Copyright 2018, WILEY-VCH Verlag GmbH & Co. KGaA, Weinheim.

(E) Transfer curve of a typical $VO_2$ electrochemical transistor, as illustrated in (D). Adapted with permission from Jo et al.[26] Copyright 2018, WILEY-VCH Verlag GmbH & Co. KGaA, Weinheim.

(F) Time-dependent conductance of a typical $VO_2$ channel (illustrated in (D)) under alternating positive and negative gate voltages. Adapted with permission from Jo et al.[26] Copyright 2018, WILEY-VCH Verlag GmbH & Co. KGaA, Weinheim.

(G) Temperature-dependent sheet resistance of a $WO_3$ film under various gate voltages. Adapted with permission from Wang et al.[60] Copyright 2017, WILEY-VCH Verlag GmbH & Co. KGaA, Weinheim.

(H) In situ $2\theta-\omega$ scans of a typical $WO_3$ film increasing the gate voltage. Adapted with permission from Wang et al.[60] Copyright 2017, WILEY-VCH Verlag GmbH & Co. KGaA, Weinheim.

(I) Depth profiles of the densities of H and D in a typical $WO_3$ film at the gate voltage of 4.5 V. Adapted with permission from Wang et al.[60] Copyright 2017, WILEY-VCH Verlag GmbH & Co. KGaA, Weinheim.

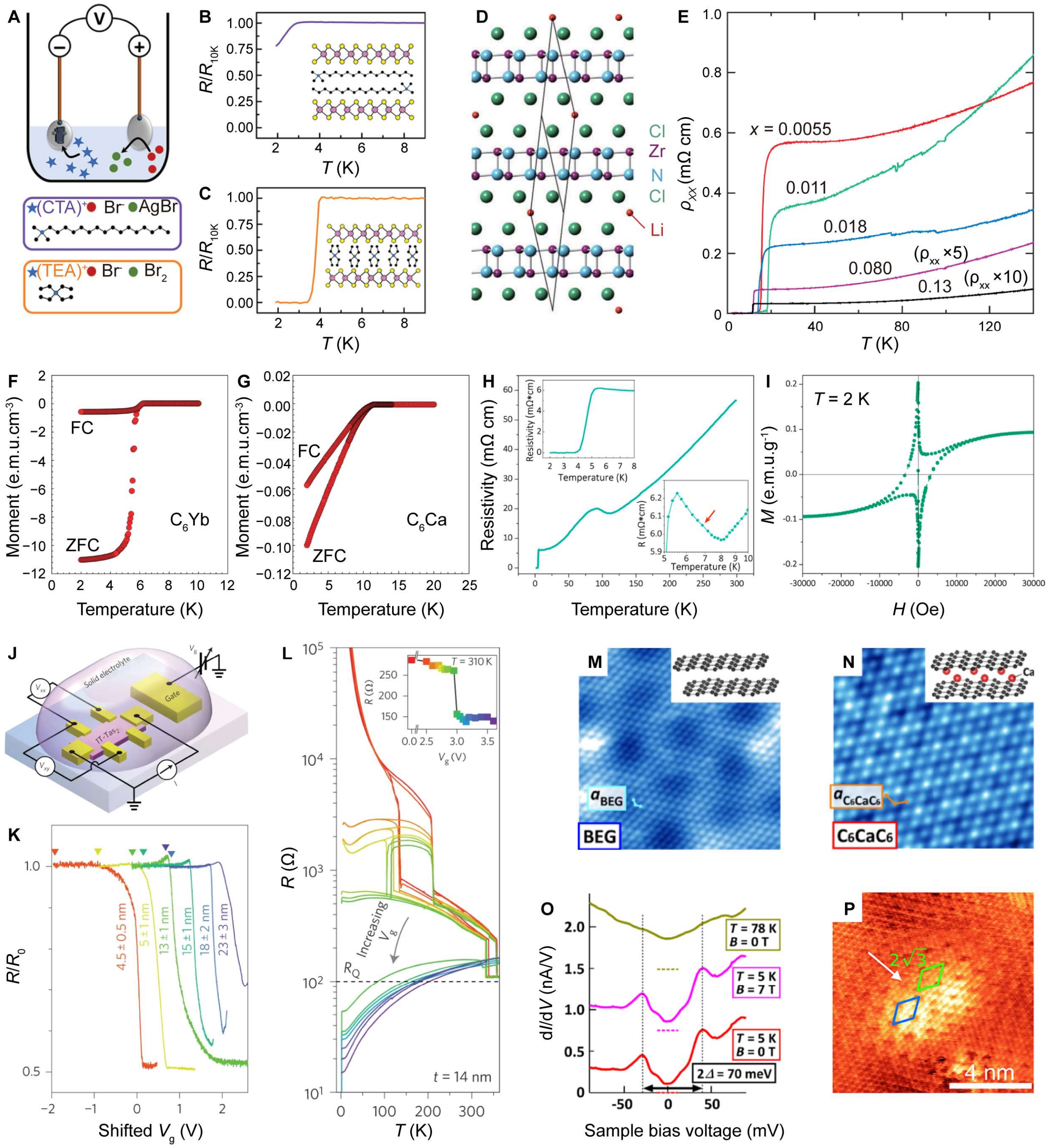

**Figure 8. Quantum phase transitions.**

(A) Schematic illustration of the experimental setup for intercalating cetyltrimethylammonium ($CTA^+$) and tetraethylammonium ($TEA^+$) ions into $MoS_2$. Adapted with permission from Pereira et al.[120] Copyright 2022, Wiley-VCH GmbH.

(B and C) Temperature-dependent resistances of $(CTA)_xMoS_2$ (B) and $(TEA)_xMoS_2$ (C). Adapted with permission from Pereira et al.[120] Copyright 2022, Wiley-VCH GmbH.

(D) The crystal structure of $Li_xZrNCl$, formed by the intercalation of $Li^+$ ions into ZrNCl. Adapted with permission from Nakagawa et al.[118] Copyright 2021, The American Association for the Advancement of Science.

(E) Resistivity-temperature curves for $Li_xZrNCl$ at different intercalation levels. Adapted with permission from Nakagawa et al.[118] Copyright 2021, The American Association for the Advancement of Science.

(F and G) Temperature-dependent magnetizations for Yb-intercalated graphite ($C_6Yb$) (F) and Ca-intercalated graphite ($C_6Ca$) (G) under zero-field-cooled (ZFC) and field-cooled (FC) conditions. Adapted with permission from Weller et al.[155] Copyright 2005, Springer Nature Limited.

(H) Temperature-dependent resistivity of a typical $SnSe_2$−Co(Cp)$_2$ nanoflake. Adapted with permission from Li et al.[50] Copyright 2017, American Chemical Society.

(I) Isothermal magnetization curve of a typical $SnSe_2$−Co(Cp)$_2$ nanoflake at 2 K. Adapted with permission from Li et al.[50] Copyright 2017, American Chemical Society.

(J) Schematic diagram of an experimental setup for intercalating $Li^+$ ions into 1T-$TaS_2$. Reproduced with permission from Yu et al.[71] Copyright 2015, Springer Nature Limited.

(K) Gate-voltage-dependent resistances for 1T-$TaS_2$ with different thicknesses. Adapted with permission from Yu et al.[71] Copyright 2015, Springer Nature Limited.

(L) Temperature-dependent resistance of a 14-nm-thick 1T-$TaS_2$ flake under different fixed gate voltages. The inset displays the gate-voltage-dependent resistance of the same flake at 310 K. Adapted with permission from Yu et al.[71] Copyright 2015, Springer Nature Limited.

(M and N) STM topographic images of pristine bilayer epitaxial graphene (BEG) (M) and Ca-intercalated bilayer graphene ($C_6CaC_6$) (N). Reproduced with permission from Shimizu et al.[139] Copyright 2015, American Physical Society.

(O) Tunneling spectra of $C_6CaC_6$ near the Fermi level under various magnetic fields and temperatures. Adapted with permission from Shimizu et al.[139] Copyright 2015, American Physical Society.

(P) STM topographic image of K-intercalated $MoS_2$ ($K_{0.4}MoS_2$) showing charge density waves of different periodicities. Adapted with permission from Bin Subhan et al.[70] Copyright 2021, The Author(s).

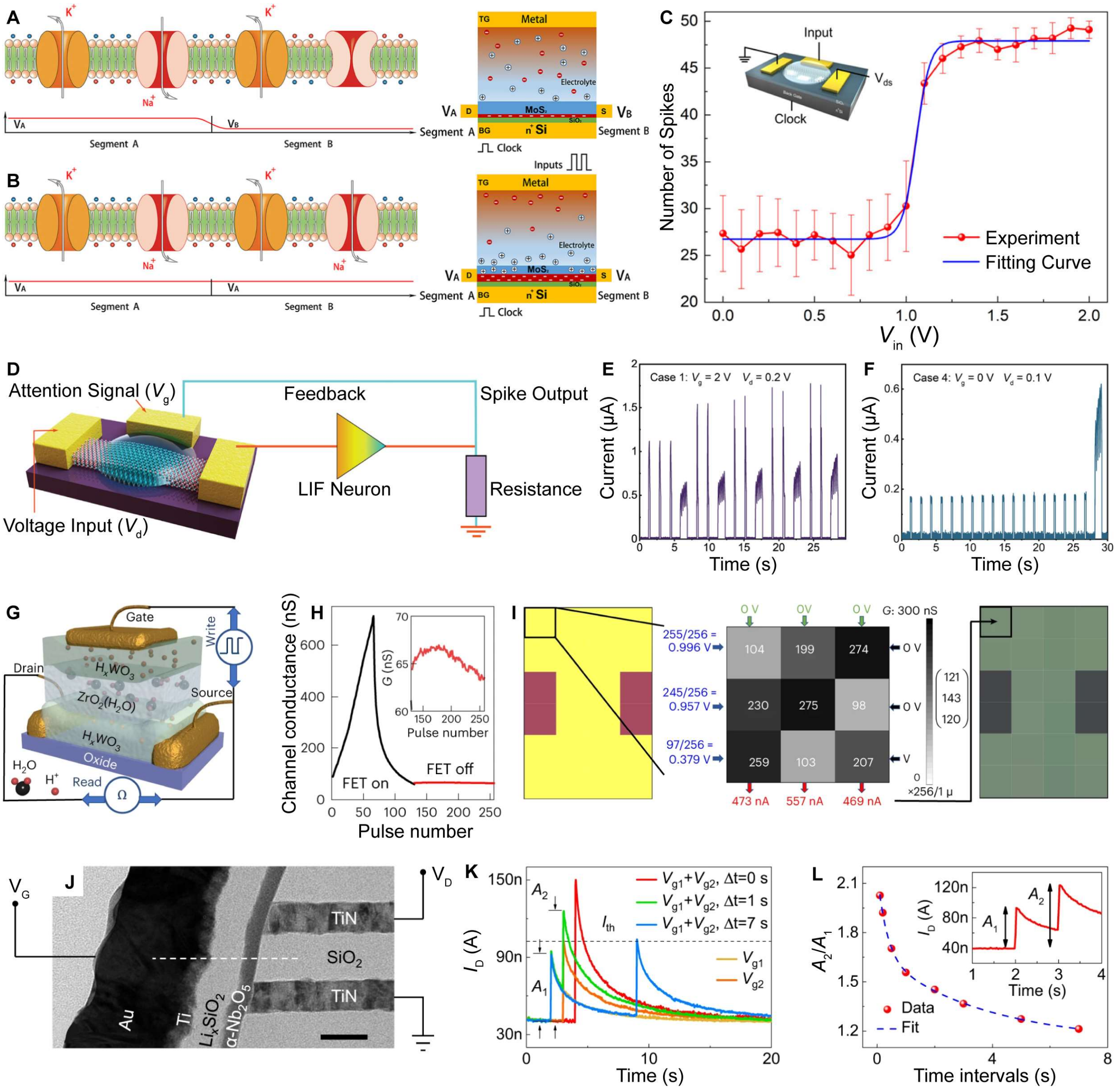

**Figure 9. Rigid devices for neuromorphic applications.**

(A) Schematic illustration of the propagation of action potential in an axon when the action potential arrives at segment A, and the membrane potential of segment B ($V_B$) is lower than the membrane potential of segment A ($V_A$), corresponding to the undoped state of the $MoS_2$ channel in a $MoS_2$ neuristor. Reproduced with permission from Bao et al.[190] Copyright 2019, American Chemical Society.

(B) Schematic illustration of the propagation of action potential in an axon when $V_B$ is increased to $V_A$ by sodium current charging, corresponding to the doped state of the $MoS_2$ channel in a $MoS_2$ neuristor. Reproduced with permission from Bao et al.[190] Copyright 2019, American Chemical Society.

(C) Evolution of the number of output spikes with the amplitude of the input pulses. Adapted with permission from Bao et al.[190] Copyright 2019, American Chemical Society.

(D) Schematic illustration of an attention computing circuit, comprising a synaptic transistor, a leaky-integrate-and-fire neuron, and a 50 KΩ monitor resistor. Adapted with permission from Liu et al.[178] Copyright 2022, The Author(s).

(E and F) The binary attention demonstrated using the circuit shown in (D): input signal “1” ($V_d$ = 0.2 V) with attention ($V_g$ = 2 V) (E) and input signal “0” ($V_d$ = 0.1 V) without attention ($V_g$ = 0 V) (F). Adapted with permission from Liu et al.[178] Copyright 2022, The Author(s).

(G) Schematic diagram of an all-inorganic $WO_3$ electrochemical transistor based on proton intercalation. Reproduced with permission from Cui et al.[77] Copyright 2023, The Author(s), under exclusive license to Springer Nature Limited.

(H) Synaptic potentiation and depression characteristics of a typical $WO_3$ electrochemical transistor (illustrated in (G)) connected with a selector transistor when the transistor is turned on or off. Adapted with permission from Cui et al.[77] Copyright 2023, The Author(s), under exclusive license to Springer Nature Limited.

(I) Demonstration of color transformation of a 4 × 6 pixel image using an array of $WO_3$ electrochemical transistors (illustrated in (G)) connected with silicon transistors. Reproduced with permission from Cui et al.[77] Copyright 2023, The Author(s), under exclusive license to Springer Nature Limited.

(J) TEM image of a vertical electrochemical transistor based on an $\alpha$-$Nb_2O_5$ channel and a $Li_xSiO_2$ electrolyte layer (scale bar: 30 nm). Adapted with permission from Xu et al.[172] Copyright 2023, The Author(s).

(K) Postsynaptic currents of a typical dual-gate transistor that is based on the vertical structure shown in (J) under individual or contemporary pulse stimulation of the two gates. Adapted with permission from Xu et al.[172] Copyright 2023, The Author(s).

(L) Dual-gate paired-pulse facilitation of the dual-gate transistor that is based on the vertical structure shown in (J). Adapted with permission from Xu et al.[172] Copyright 2023, The Author(s).

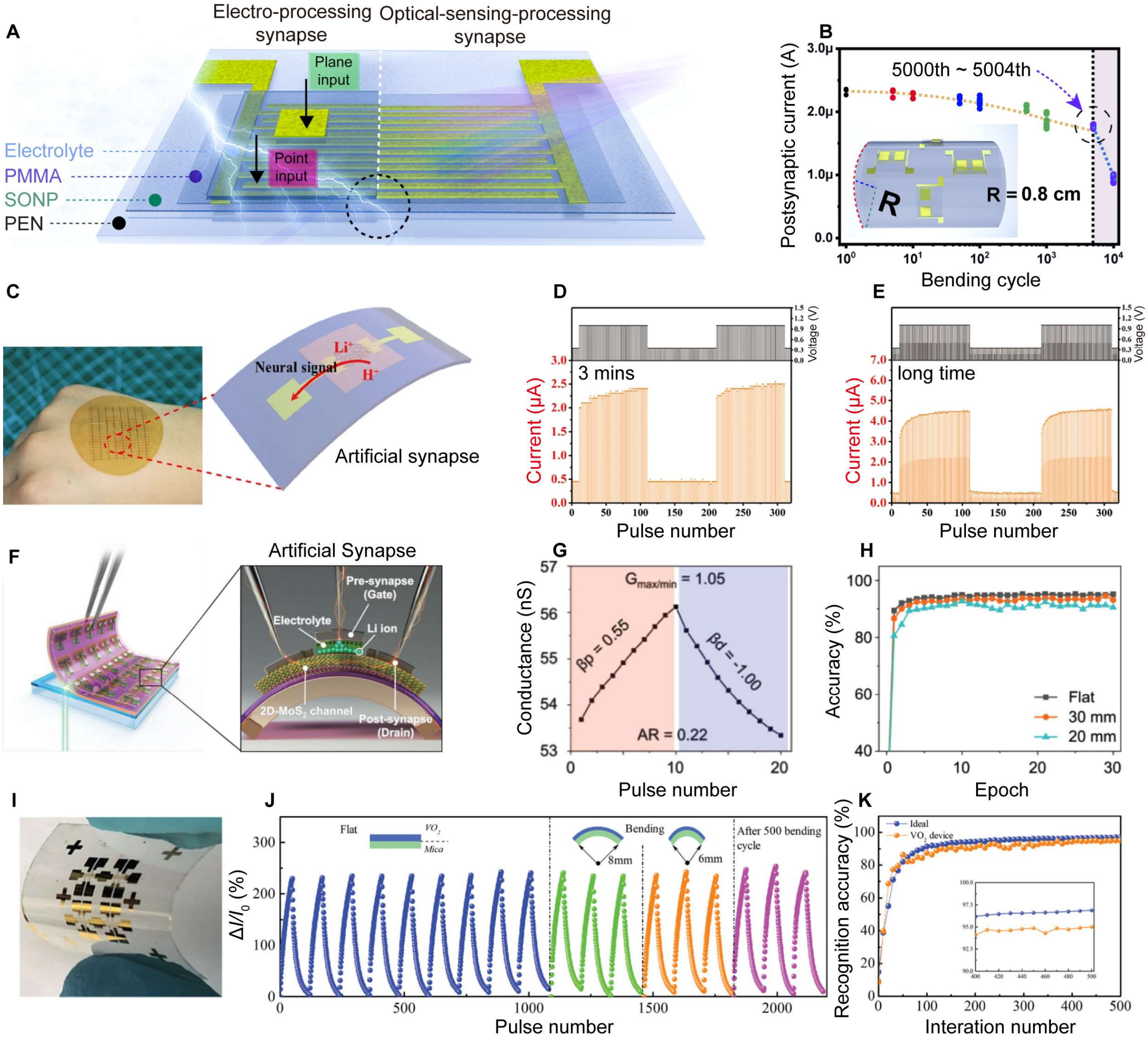

**Figure 10. Flexible devices for neuromorphic applications.**

(A) Schematic illustration of a $SnO_2$ nanoparticles (SONPs)-based flexible electrochemical transistor, which can be used to emulate synaptic behaviors under the stimulation of either electrical or optical signals. Adapted with permission from Wei et al.[93] Copyright 2020, Elsevier Ltd.

(B) The relationship between postsynaptic currents and bending cycles of a typical device illustrated in (A). Adapted with permission from Wei et al.[93] Copyright 2020, Elsevier Ltd.

(C) Optical image of an array of flexible two-terminal neuromorphic devices based on semiconducting single-wall carbon nanotubes intercalated with protons and $Li^+$ ions, as illustrated in the schematic image. Adapted with permission from Lu et al.[192] Copyright 2020, The Author(s).

(D) Emulation of the complete recovery process of the nerve after mild stimuli using a typical device illustrated in (C). Adapted with permission from Lu et al.[192] Copyright 2020, The Author(s).

(E) Emulation of the incomplete recovery of the nerve after severe stimuli using a typical device illustrated in (C). Adapted with permission from Lu et al.[192] Copyright 2020, The Author(s).

(F) Schematic diagram of a $MoS_2$-based flexible neuromorphic device. Adapted with permission from Hwang et al.[195] Copyright 2023, The Author(s).

(G) Potentiation/depression characteristics of a typical neuromorphic device illustrated in (F). Adapted with permission from Hwang et al.[195] Copyright 2023, The Author(s).

(H) Simulated recognition accuracy of handwritten sets of the network based on the neuromorphic device illustrated in (F) bent at different radii. Adapted with permission from Hwang et al.[195] Copyright 2023, The Author(s).

(I) Optical image of a $VO_2$-based flexible Mott synaptic transistor. Reproduced with permission from Deng et al.[171] Copyright 2021, Wiley-VCH GmbH.

(J) Endurance properties of a typical neuromorphic device illustrated in (I) bent at different radii and after 500-bending cycles at a bending radius of 6 mm. Adapted with permission from Deng et al.[171] Copyright 2021, Wiley-VCH GmbH.

(K) Comparison of the simulated recognition accuracy between an ideal synaptic device and a $VO_2$-based synaptic device illustrated in (I). Adapted with permission from Deng et al.[171] Copyright 2021, Wiley-VCH GmbH.

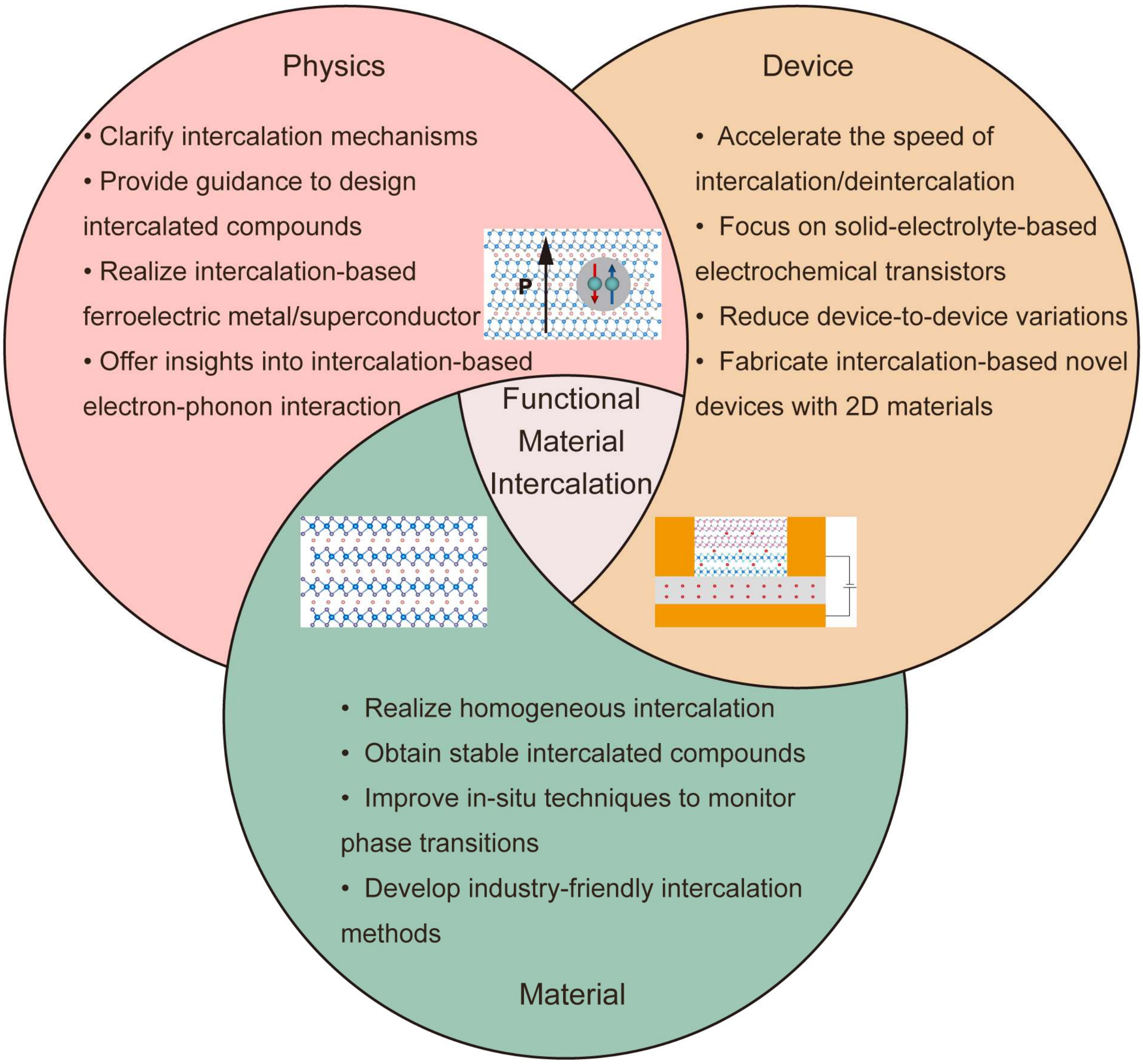

**Figure 11. Possible future research lines for functional material intercalation.**